\documentclass[a4paper,11pt]{article}
\pdfoutput=1 

\usepackage{jcappub} 
                     
\usepackage{txfonts} 
\usepackage{tensor}
\usepackage{mathtools}
\usepackage{cleveref}
\usepackage{tikz,pgfplots}
\usepackage{upgreek}
\usepackage{amssymb}

\newcommand{\ltsima}{$\; \buildrel < \over \sim \;$}
\newcommand{\lsim}{\lower.5ex\hbox{\ltsima}}
\newcommand{\gtsima}{$\; \buildrel > \over \sim \;$}
\newcommand{\gsim}{\lower.5ex\hbox{\gtsima}}

\newcommand{\chip}{{\chi^\prime}}

\newcommand{\review}[1]{{\color{black}{#1}}}
\newcommand{\newreview}[1]{{\color{black}{#1}}}
\newcommand{\reviewnew}[1]{{\color{black}{#1}}}

\title{\boldmath Calibrating baryonic feedback with weak lensing and fast radio bursts}

\author[a,b]{Robert Reischke,}
\author[a]{Dennis Neumann,}
\author[a]{Klara Antonia Bertmann,}
\author[c,d]{Steffen Hagstotz}
\author[a]{and Hendrik Hildebrandt}

\affiliation[a]{Ruhr University Bochum, Faculty of Physics and Astronomy, Astronomical Institute (AIRUB), German Centre for Cosmological Lensing, 44780 Bochum, Germany}
\affiliation[b]{Argelander-Institut für Astronomie, Universität Bonn, Auf dem Hügel 71, D-53121 Bonn, Germany}
\affiliation[c]{Universitäts-Sternwarte, Fakultät für Physik, Ludwig-Maximilians Universität München,
Scheinerstraße 1, D-81679 München, Germany}
 \affiliation[d]{
Excellence Cluster ORIGINS, Boltzmannstraße 2, D-85748 Garching, Germany}

\emailAdd{reischke@posteo.net}

\abstract{One of the key limitations of large-scale structure surveys of the current and future generation, such as Euclid, LSST-Rubin or Roman, is the influence of feedback processes on the distribution of matter in the Universe. This effect, called baryonic feedback, modifies the matter power spectrum on non-linear scales much stronger than any cosmological parameter of interest. Constraining these modifications is therefore key to unlocking the full potential of the upcoming surveys, and we propose to do so with the help of Fast Radio Bursts (FRBs). FRBs are short, astrophysical radio transients of extragalactic origin. Their burst signal is dispersed by the free electrons in the large-scale structure, leading to delayed arrival times at different frequencies characterised by the dispersion measure (DM). Since the dispersion measure is sensitive to the integrated line-of-sight electron density, it is a direct probe of the baryonic content of the Universe.
We investigate how FRBs can break the degeneracies between cosmological and feedback parameters by correlating the observed Dispersion Measure with the weak gravitational lensing signal of a Euclid-like survey. In particular, we use a simple one-parameter model controlling baryonic feedback, but we expect similar findings for more complex models. \reviewnew{Within this model, we find that  $\sim 5\times 10^4$ FRBs are sufficient to constrain the baryonic feedback significantly better than cosmic shear alone, tightening the constraints considerably (roughly by a factor of five). We also expect a 1.5-fold improvement in the sum of neutrino masses.}}
\pgfplotsset{compat=1.18}
\begin{document}
\maketitle
\flushbottom

\section{Introduction}
Cosmic shear, the weak gravitational lensing effect imprinted on distant galaxies by the large-scale structure (LSS), is one of the primary science goals for the upcoming stage-4 surveys Euclid\footnote{\href{https://www.euclid-ec.org/}{https://www.euclid-ec.org/}}, Rubin-LSST\footnote{\href{https://www.lsst.org/}{https://www.lsst.org/}} and the Roman telescope\footnote{\href{https://roman.gsfc.nasa.gov/}{https://roman.gsfc.nasa.gov/}}. The foundations for these missions has been laid down by their predecessors: the Kilo-Degree Survey \citep[KiDS]{hildebrandt_kids-450_2017,asgari_kids-1000_2021}\footnote{\href{https://kids.strw.leidenuniv.nl/}{https://kids.strw.leidenuniv.nl/}}, the Dark Energy Survey \citep[DES]{abbott_dark_2018,amon_dark_2022}\footnote{\href{https://www.darkenergysurvey.org/}{https://www.darkenergysurvey.org/}} and the Subaru Hyper Suprime-Cam \citep[HSC]{ hamana_cosmological_2020, 2023PhRvD.108l3519D}\footnote{\href{https://hsc.mtk.nao.ac.jp/ssp/}{https://hsc.mtk.nao.ac.jp/ssp/}}. 

The stage-4 surveys in particular target the highly non-linear regime of the matter distribution. Even though there is a wealth of information to be gathered at $k>1\,\mathrm{Mpc}^{-1}h$, baryonic feedback (BF) processes imprint significant theoretical uncertainties on the total matter power spectrum. While the distribution of collisionless cold dark matter is well understood, at least on the level of numerical simulations, baryons remain the cause of significant uncertainties. The displacement of baryons is due to nonlinear effects sourced by star formation, Supernovae, Active Galactic Nuclei (AGN) or cooling \citep[see e.g.][for reviews]{somerville_physical_2015,vogelsberger_cosmological_2020}. In turn, the displaced baryons affect the total matter power spectrum, reducing power on intermediate scales due to outflows and energy injection by e.g. AGN, and increasing power on very small scales due to cooling. As these effects span a multitude of scales and are highly non-linear, they are almost impossible to model analytically and are thus extracted from hydrodynamical simulations, which include a model with a fixed set of parameters to control the feedback strength. The main challenge arises from the fact that feedback processes originate in physics, which is far below the numerical simulation's resolution. Each simulation, therefore, has sub-grid models implemented. However, different models can lead to significantly different predictions for the matter clustering statistics at small scales \citep{chisari_impact_2018,vandaalen_exploring_2020}. For example, in the \texttt{CAMELS} simulation suite \citep{villaescusa-navarro_camels_2021}, two very distinct sub-grid models are implemented. However, the differences are largely due to higher gas temperature in the intergalactic medium \citep{villaescusa-navarro_camels_2021}. Another suite of simulations is the \texttt{BAHAMAS} simulations \citep{mccarthy_bahamas_2017} whose power spectra have been calibrated to a hydrodynamical version of the halo model \citep{mead_hydrodynamical_2020}, designed explicitly for weak lensing cross-correlation studies with the thermal Sunyaev-Zel'dovich (tSZ) effect \citep{troster_joint_2022} caused by the thermal velocity of hot electrons in galaxy clusters. 

Any independent measurement of the baryon distribution can help constrain feedback models independently. It was shown in \citep{troster_joint_2022} that, e.g. the tSZ effect has the potential to partially calibrate the feedback strength and break degeneracies (see also a discussion \citep{nicola_breaking_2022}). \review{Another related possibility to measure the electron distribution uses not thermal, but the bulk velocities of galaxy clusters for the kinetic SZ (kSZ) effect \citep{sunyaev_observations_1972,sunyaev_microwave_1980}. This was done by various studies using data from the Atacama Cosmology Telescope \citep{Hand:2012ui,PhysRevD.93.082002}, from Planck \citep{Planck:2015ywj} and SPT \citep{DES:2016umt} and can also break degeneracies between feedback models \citep{PhysRevD.103.063513, 2021PhRvD.103f3514A}. \newreview{These measurements start to be able to discriminate between different feedback models. It was for example shown in \cite{2024arXiv240707152H} prefer stronger feedback which was supported recently as well by \cite{2024arXiv241019905M}.}
At the moment, the tSZ and kSZ effects are the primal ways to measure the electron distribution in the Universe. 
It is therefore important to add additional probes for $(i)$ more constraining power and $(ii)$ different systematic effects in the measurements.}

An alternative probe of the electron distribution are
fast radio bursts (FRBs). FRBs are short transients, lasting typically a few milliseconds, with frequencies ranging from $\sim 100$ MHz to several GHz. Since the radio signal is travelling through the ionised intergalactic medium (IGM), each frequency of the pulse experiences a delay characterised by the dispersion measure (DM) proportional to the integrated free electron density along the line-of-sight \citep[e.g.][]{thornton_population_2013, petroff_real-time_2015, connor_non-cosmological_2016, champion_five_2016,chatterjee_direct_2017}. The mechanism responsible for the radio emission is currently unknown, but the isotropic occurrence of the events detected so far shows no alignment with the Milky Way disk, and the measured DM for most events is very large, requiring an extragalactic origin. By now, a subset of events has been localised to galaxies up to redshifts $z \sim 1$. In general, the total DM associated with an extragalactic FRB event consists of contributions from the host galaxy, the Milky Way and the diffuse electrons in the large-scale structure (LSS). Several authors, therefore, proposed to use the DM inferred from FRBs as a cosmological probe using either the averaged signal \citep[e.g.][]{zhou_fast_2014,walters_future_2018,hagstotz_new_2022,macquart_census_2020,wu_8_2022,james_measurement_2022,reischke_consistent_2023,reischke_covariance_2023} or the statistics of DM fluctuations \citep[e.g.][]{masui_dispersion_2015,shirasaki_large-scale_2017,rafiei-ravandi_chimefrb_2021,bhattacharya_fast_2021,takahashi_statistical_2021,reischke_probing_2021, reischke_consistent_2022}.

Currently, roughly $\sim 1000$ FRBs are detected, but estimated rates of $\sim 1000$ events/night/sky allow surveys such as CHIME, UTMOST, HIRAX, ASKAP, CHORD, DSA-2000 or SKA \citep{johnston_science_2008, 2020PASA...37....7S, caleb_fast_2016,newburgh_hirax_2016,2019BAAS...51g.255H, 2019clrp.2020...28V,2020PASA...37....7S} to provide at least $\sim 10^4$ FRBs per decade. Although FRBs do not show spectral features that allow for accurate redshift estimation, the accumulated DM can be translated into a noisy distance estimate \citep{masui_dispersion_2015}. Since FRBs are observed with $\mathrm{DM} \geq 1000\;\mathrm{pc}\;\mathrm{cm}^{-3}$ and models of the Milky Way suggest $\mathrm{DM} \leq50\;\mathrm{pc}\;\mathrm{cm}^{-3}$ due to gas in our Galaxy for most directions on the sky \citep{yao_new_2017}, the majority of the DM signal originates from the diffuse electrons in the IGM, which follows the total matter distribution on large scales. Although the host galaxy contribution is still under debate, its magnitude is expected to be comparable to that of the Milky Way and, as a result, usually smaller than the contribution from the LSS. While the host dispersion can reach several hundreds $\mathrm{pc}\;\mathrm{cm}^{-3}$ in rare cases, particularly when the FRB is originates in the trailing edge of the host galaxy (as seen by the observer), one typically finds that $\mathrm{DM}_\mathrm{LSS} > \mathrm{DM}_\mathrm{host}$ for FRBs at redshifts $z\sim 1$.
The dispersion of FRBs tests, similar to cosmic shear, is an integrated quantity along the line-of-sight. However, since host galaxies typically contribute less to the DM than the LSS, FRBs are way less affected by shot noise compared to cosmic shear. Combining FRBs and cosmic shear will accordingly probe similar structures, but testing the baryon distribution and the total matter distribution respectively, enabling strong constraints on the different clustering behaviour of both components. It should be noted that the cosmic shear signal has a vanishing expectation value, unlike the DM of FRBs. However, since we are interested in the statistical properties of the DM, this is not particularly important for our study.

In this paper, we will hence investigate the potential gains from combining a stage-4 cosmic shear measurement with the statistical properties of the DM measured from an FRB sample, which should be available on the same timescale as cosmic shear data from Euclid and Rubin-LSST. To this end, we simulate mock data for these kinds of surveys and carry out a Fisher and MCMC analysis to investigate the impact of adding FRBs in a cosmological analysis with a particular focus on the baryonic feedback and on the cosmological parameters that are most affected by feedback uncertainties, such as the sum of neutrino masses. The paper is structured as follows:
In \Cref{sec:probes}, we summarise the LSS structure probes and describe the methodology for our analysis. \Cref{sec:results} is devoted to the discussion of the results and a comparison with other probes of the electron distribution. Lastly, we conclude in \Cref{sec:conclusion} and discuss potential ways forward.

\section{Large-Scale-Structure Probes}
\label{sec:probes}
In this section, we summarise the two probes used to test the matter distribution: cosmic shear and FRB dispersion measure. We introduce cosmic shear in \Cref{sec:cosmic_shear} and DM correlations in \Cref{sec:modelling_frb}, 
which are sensitive to the full matter and electron distribution, respectively, and can hence be used to break degeneracies between baryonic feedback and cosmological parameters.

\subsection{Cosmic Shear}
\label{sec:cosmic_shear}
Cosmic shear is the gravitational lensing effect of the LSS on an ensemble of background sources (galaxies) and is thus sensitive to the projected matter distribution along the line-of-sight.
The cosmological information of gravitational lensing by the LSS is contained in the traceless part of the cosmic shear tensor:
\begin{equation}
    \boldsymbol{\gamma}_i = 2\int\mathrm{d}\chi W_{\gamma_i}(\chi)\eth\eth \Phi (\hat{\boldsymbol{x}},\chi)\;,
\end{equation}
in the absence of anisotropic stress. Here $\gamma_i$ denotes the shear in the $i$-th tomographic bin, $W_{\gamma_i}$ is a geometrical weighting function defined below, $\Phi$ are the scalar metric perturbations, $\chi$ the comoving distance and $\eth$ is the \textit{eth}-derivative for a spin-2 field. Using Poisson's equation, one can relate the metric perturbations to the density contrast, yielding the following expression for the angular power spectrum:
\begin{align}\label{eq:lensing_nonlimber}
C^{\gamma_i\gamma_j}_\ell = 
\frac{(\ell + 2)!}{(\ell -2)!}\frac{2}{\uppi}\int\mathrm{d}\chi_1 \int\mathrm{d}\chi_2\int k^2\mathrm{d}k\; K_i(\chi_1) K_j(\chi_2) \frac{j_\ell(\chi_1 k)}{(k\chi_1)^2}\frac{j_\ell(\chi_2k)}{(k\chi_2)^2} \sqrt{P_{\delta}(k,\chi_1)P_{\delta}(k,\chi_2)}\;. 
\end{align}
Approximating the Bessel function by a Dirac distribution
\citep{limber_analysis_1954,loverde_extended_2008}, the integrations over $\chi_{1,2}$ can be carried out to find the commonly used expression for the angular power spectrum:
\begin{equation}
\label{eq:limber}
    C^{\gamma_i\gamma_j}_\ell =  \int_0^{\chi^{\mathstrut}_\mathrm{H}}\frac{\mathrm{d}\chi}{\chi^2} W^{(i)}_\gamma(\chi)W^{(j)}_\gamma(\chi) P_\delta\left(\frac{\ell + 1/2}{\chi},\chi\right)\;,
\end{equation}
which will be accurate enough for cosmic shear even for the next generation of surveys \citep{leonard_n5k_2022} due to the broad lensing kernel. Here $P_\delta$ is the matter power spectrum, for which we use the emulated spectrum from $\texttt{HMcode}$\citep{mead_accurate_2015}. $W^{(i)}_\gamma (\chi)$ is the lensing weight of the $i$-th tomographic bin as given by:
\begin{equation}
\label{eq:weight}
    W^{(i)}_\gamma(\chi) =\frac{3\Omega_\mathrm{m0}}{2\chi^{2}_\mathrm{H}}\frac{\chi}{a(\chi)}\int_\chi^{\chi^{\mathstrut}_\mathrm{H}}\mathrm{d}\chip n^{(i)}_\mathrm{s}(\chip)\frac{\chip-\chi}{\chip}\;.
\end{equation}
Here $a$ is the scale factor, $\Omega_{\mathrm{m}0}$ the matter density parameter today, $\chi^{\mathstrut}_\mathrm{H}$ the Hubble radius and $n^{(i)}_\mathrm{s}$ is the distribution of sources. 

Lastly, due to the finite number of background galaxies, observed spectra obtain a shape-noise contribution:
\begin{equation}
\label{eq:observed_lensing_spectra}
    C^{\gamma_i\gamma_j}_\ell\;\to\; C^{\gamma_i\gamma_j}_\ell + \frac{\sigma_{\epsilon,i}^2}{2\bar{n}_i}\delta^\mathrm{K}_{ij}\;,
\end{equation}
with the total ellipticity dispersion $\sigma_{\epsilon,i}^2$ and the average number of sources $\bar{n}_i$ in the $i$-th tomographic bin. $\delta^\mathrm{K}_{ij}$ is the Kronecker delta, and ensures that the intrinsic shapes of different redshift bins are uncorrelated.

\renewcommand{\arraystretch}{1.5}
\begin{table}
\footnotesize
    \centering
    \begin{tabular}{ccccc}
     Survey &  area $f_\mathrm{sky}$ & number of sources & intrinsic noise $\sigma$ & tomographic bins\\
     \hline\hline
     FRB & 0.7  & $5\times 10^4$ & $50\; \mathrm{pc}\;\mathrm{cm}^{-3}$ & 1\\
     Euclid & 0.3  & $30\;\mathrm{arcmin}^{-2}$ & 0.3 & 10\\ 
    \end{tabular}
    \caption{Survey settings assumed in the analysis. The number of sources for the cosmic shear surveys is an average density over all tomographic bins. The actual number in each tomographic bin may differ slightly. The same holds for the ellipticity dispersion; for more details, see \citep{blanchard_euclid_2020} and \citep{joachimi_kids-1000_2020}. \review{Note that we give the ellipticity dispersion of a single component for cosmic shear and that the intrinsic noise is the averaged noise over the whole FRB sample.}}
    \label{tab:survey_settings}
\end{table}

\subsection{FRB statistics}
\label{sec:modelling_frb}
FRB pulses undergo dispersion while travelling through the ionised IGM, causing a frequency-dependent (proportional to $\nu^{-2}$) offset of arrival times. The corresponding time delay measured $\delta t(\hat{\boldsymbol{x}},z)$ for an FRB at redshift $z$ in direction $\hat{\boldsymbol{x}}$ proportional to the observed dispersion measure: $\delta t(\hat{\boldsymbol{x}},z) = \mathrm{DM}_\mathrm{tot}(\hat{\boldsymbol{x}}, z) \nu^{-2}$.
The observed DM can be broken up into different components:
\begin{equation}
\label{eq:dispersion_measure_contributions}
    \mathrm{DM}_\mathrm{tot}(\hat{\boldsymbol{x}},z) = \mathrm{DM}_\mathrm{LSS}(\hat{\boldsymbol{x}},z) + \mathrm{DM}_\mathrm{MW}(\hat{\boldsymbol{x}}) + \mathrm{DM}_\mathrm{host}(z)\;,
\end{equation}
where $\mathrm{DM}_\mathrm{LSS}(\hat{\boldsymbol{x}},z)$ is the DM caused by the electron distribution in the LSS, while $\mathrm{DM}_\mathrm{MW}(\hat{\boldsymbol{x}})$ and $\mathrm{DM}_\mathrm{host}(z)$ describe the contributions from the Milky Way and the host galaxy, respectively. Here, we made the dependence on redshift and direction explicit. Note that the rest-frame DM of the host, $\mathrm{DM}_\mathrm{host,rf}$, is observed as $\mathrm{DM}_\mathrm{host}(z) = (1+z)^{-1}\mathrm{DM}_\mathrm{host,rf}$.

 Current models of the galactic electron distribution predict ${\mathrm{DM}_\mathrm{MW} \sim 60 \, \mathrm{pc}\,\mathrm{cm}^{-3}}$ over most of the sky \citep{yao_new_2017,platts_data-driven_2020}. Since the galactic electron distribution also depends on the direction $\hat{\boldsymbol{x}}$, it is in principle a large contaminant for DM correlation measurement. Here, however, we assume that it can be modelled accurately enough to be subtracted from the observed DM without leaving a significant imprint on the measured DM correlations. Naively, one can expect $\mathrm{DM}_\mathrm{host}(z)$ to be very similar to $\mathrm{DM}_\mathrm{MW}$. However, in practice, it depends strongly on the host galaxy's properties and is thus treated as a free parameter. Since it does not correlate with the LSS, it will not leave an imprint on the DM correlation but rather act as a noise term, as we will discuss later. Finally, the LSS contribution to the DM is given by \citep{reischke_probing_2021} 
\begin{equation}
\label{eq:dispersion_measure_general}
    \mathrm{DM}_\mathrm{LSS}(\hat{\boldsymbol{x}},z) = \chi^{\mathstrut}_\mathrm{H}\int_0^z \mathrm{d}z^\prime \, n_\mathrm{e}({\boldsymbol{x}},z^\prime) \, \frac{1+z^\prime}{E(z^\prime)} \,  \;,
\end{equation}
where $E(z)$ is the expansion function and $n_e$  the electron density, which depends on the local matter over-density, $\delta_\mathrm{m}(\boldsymbol{x},z)$ and the electron bias $b_\mathrm{e}({\boldsymbol{x}},z)$:
\begin{equation}
\label{eq:electron_number_density}
    n_\mathrm{e}({\boldsymbol{x}},z) = \frac{\bar\rho_\mathrm{b}(z)}{m_\mathrm{p}} F(z) \left[1+b_\mathrm{e}({\boldsymbol{x}},z) \, \delta_\mathrm{m}({\boldsymbol{x}},z)\right] \; .
\end{equation}
Here we used the mean baryon density $\bar\rho_\mathrm{b}(z)$, the proton mass $m_\mathrm{p}$ and the mass fraction $F(z)$ of electrons in the IGM. The latter can be expressed as:
\begin{equation}
\label{eq:ionization_fraction_of_the_igm}
    F(z) = f_\mathrm{IGM}(z)[Y_\mathrm{H}X_{\mathrm{e},\mathrm{H}}(z) + Y_\mathrm{He}X_{\mathrm{e},\mathrm{He}}(z)]\;,
\end{equation}
with the mass fraction of hydrogen and helium $Y_\mathrm{H} = 0.75$ and $Y_\mathrm{He} = 0.25$ respectively, as well as their ionization fractions $X_{\mathrm{e},\mathrm{H}}(z)$ and $X_{\mathrm{e},\mathrm{He}}(z)$. Since both hydrogen and helium are both fully ionised for $z\lsim 3$ \citep{meiksin_physics_2009,becker_detection_2011} we set $X_{\mathrm{e},\mathrm{H}} =X_{\mathrm{e},\mathrm{He}} = 1$. It should be noted that for the survey settings considered here, the majority of the signal arises from $z<3$, so the last assumption is indeed well justified. The fraction of baryons in the IGM, $f_\mathrm{IGM}(z)$, has a slight redshift dependence \citep{shull_baryon_2012}, with $10\%\; (20\%)$ locked up in galaxies at $z \gsim 1.5 \; (\lsim 0.4)$. Rewriting \cref{eq:dispersion_measure_general} leaves us with the following expression:
\begin{equation}
\label{eq:dispersion_measure_specific}
  \!\!  \mathrm{DM}_\mathrm{LSS}(\hat{\boldsymbol{x}},z) = \mathcal{A}\!\int_0^z\!\!\! \mathrm{d}z^\prime \frac{1+z^\prime}{E(z^\prime)}F(z^\prime)[1\!+\!b_\mathrm{e}({\boldsymbol{x}},z^\prime)\delta_\mathrm{m}({\boldsymbol{x}},z^\prime)]\;,
\end{equation}
with the amplitude
\begin{equation}
\label{eq:DM_amplitude}
    \mathcal{A} \equiv \frac{3H_0^2\Omega_{\mathrm{b}0}\chi^{\mathstrut}_\mathrm{H}}{8\uppi G m_\mathrm{p}}\;.
\end{equation}
A very intriguing aspect of FRBs is that the cosmological contribution is typically larger than the contribution from the host or the Milky Way. This is in significant contrast to cosmic shear, where the ellipticity imprinted by gravitational lensing is a small perturbation of the intrinsic ellipticity. Therefore, far fewer FRBs are required to detect DM correlations with a high significance than the number of galaxies in weak lensing studies.
Following \citep{reischke_probing_2021}, we rewrite the LSS contribution to the DM as a background contribution and a perturbation:
\begin{equation}
\label{eq:fluctuation_definition}
     \mathrm{DM}_\mathrm{LSS}(\hat{\boldsymbol{x}},z) = \langle  \mathrm{DM}_\mathrm{LSS}\rangle (z) + \mathcal{D}(\hat{\boldsymbol{x}},z)\;,
\end{equation}
where
$\mathcal{D}(\hat{\boldsymbol{x}},z)$ is the effective DM induced by the fluctuations in the LSS. In complete analogy to cosmic shear, we assume a source-redshift distribution $n_\mathrm{FRB}(z)$. \reviewnew{As default, we will assume that the redshifts of the FRBs are known and come back to this issue in \Cref{app:redshifts}.}
By rearranging integration limits, one finds
\begin{equation}
\label{eq:averaged_dispersion_measure_fluctuation}
    \mathcal{D}(\hat{\boldsymbol{x}}) = \int_0^{\chi^{\mathstrut}_\mathrm{H}}\!\!\mathrm{d}\chi\; W_\mathcal{D}(\chi)\delta_\mathrm{e}
    \big(\hat{\boldsymbol{x}},z(\chi)\big) \;,
\end{equation}
with the averaged weighting function
\begin{equation}
\label{eq:averaged_dispersion_measure_fluctuation_weighting_function}
    W_\mathcal{D}(\chi) = W(\chi)\int_\chi^{\chi^{\mathstrut}_\mathrm{H}} \mathrm{d}\chi^\prime n_\mathrm{FRB}(\chi^\prime) \; ,
\end{equation}
and $W(\chi)$ being defined via \cref{eq:dispersion_measure_specific}:
\begin{equation}
\label{eq:averaged_dispersion_measure_fluctuation_weighting_function_inner}
    W(\chi) = \mathcal{A}\frac{F \big(z(\chi) \big) \big(1+z(\chi) \big)}{E\big(z(\chi)\big)}\left|\frac{\mathrm{d}z}{\mathrm{d}\chi}\right|\;.
\end{equation}
In complete analogy to cosmic shear, the angular power spectrum for the DM fluctuations $\mathcal{D}$ is given by:
\begin{equation}
\label{eq:dispersion_measure_angular_power_spectrum}
    C^{\mathcal{D}\mathcal{D}}(\ell) =  \;\frac{2}{\uppi}\int\mathrm{d}\chi_1 \int\mathrm{d}\chi_2\int k^2\mathrm{d}k\;W_\mathcal{D}(\chi_1)  W_\mathcal{D}(\chi_2)\sqrt{P_{\mathrm{ee}}(k,\chi_1)P_{\mathrm{ee}}(k,\chi_2)}j_{\ell}(k\chi_1)j_{\ell}(k\chi_2)\;.
\end{equation}
As discussed in \citep{reischke_probing_2021}, the Limber approximation is suitable for DM correlations as well. Hence, we again use \cref{eq:limber}. When cross-correlating cosmic shear and the DM, we replace one of the weight functions in \Cref{eq:limber} with $W_\mathcal{D}$ and the matter power spectrum by the electron-matter cross power spectrum $P_{\delta\mathrm{e}}$.

\begin{figure}
    \centering
    \includegraphics[width = .48\textwidth]{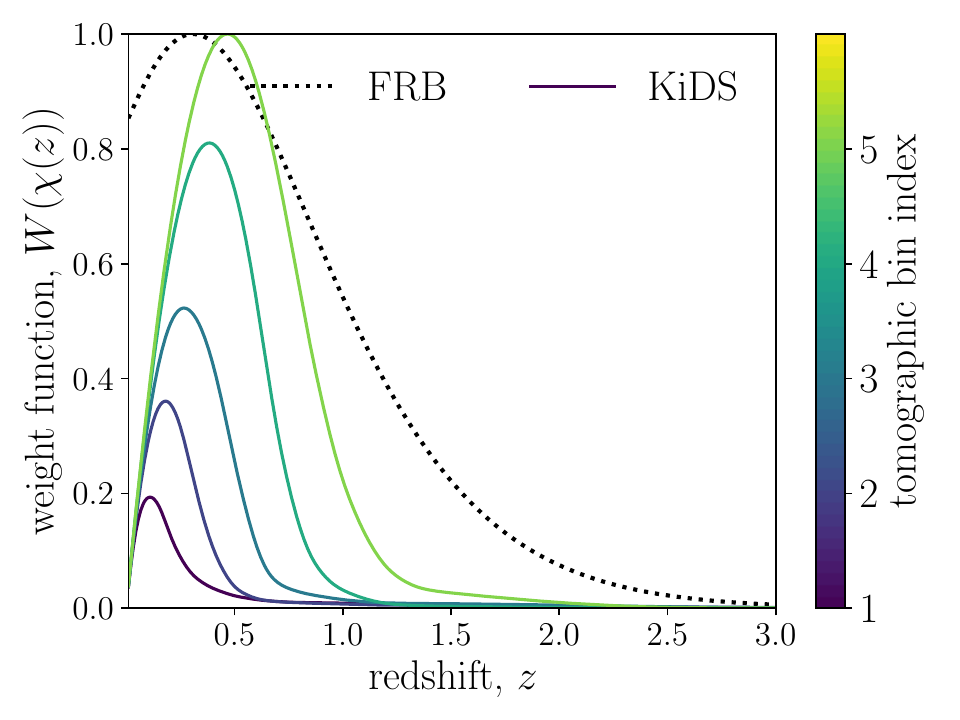}
    \includegraphics[width = .48\textwidth]{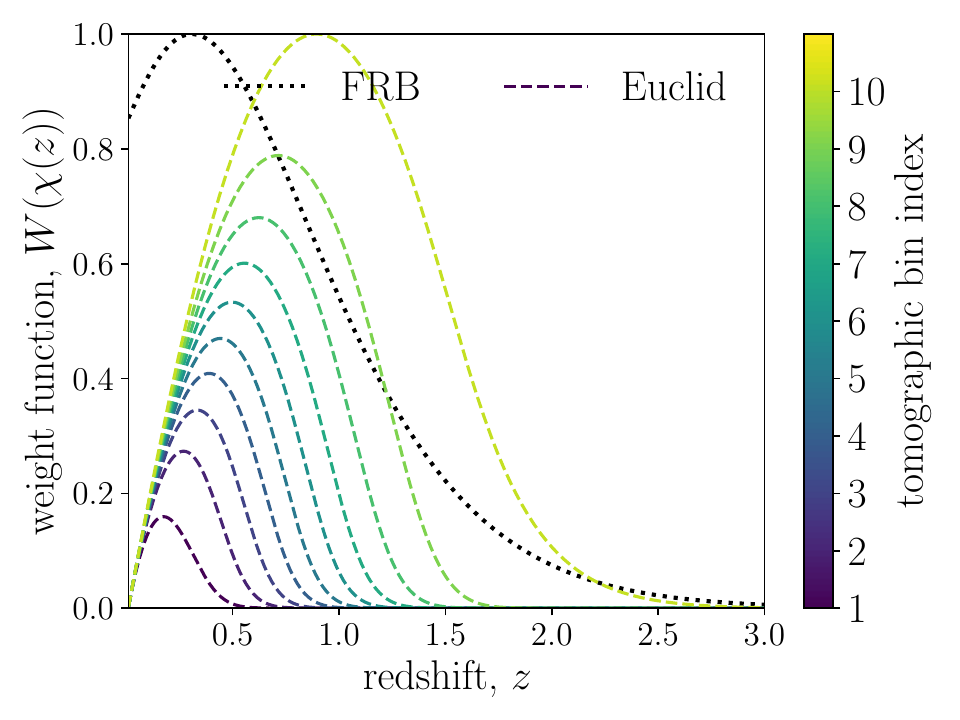}
    \caption{Weighting function for FRBs (dotted black line) and for cosmic shear with the KiDS redshift distribution (left plot) or Euclid's redshift distribution (right plot). The colour bar shows the tomographic bin index. All weighting functions have been normalised to the peak value. For cosmic shear, the peak of the last tomographic bin was used.}
    \label{fig:weights}
\end{figure}

\reviewnew{Lastly, each observed field will have an intrinsic field noise component. This arises because the cosmological field under consideration produces random fluctuations for each observed line-of-sight, introducing an irreducible scatter due to cosmic variance. In other words, even if we subtract the mean contribution, the fluctuations in the field are a single realisation of the field at that point\footnote{This point is discussed formally, for example, in \cite{2025OJAp....8E..57F}.}. 
For cosmic shear, this is negligible since the intrinsic ellipticity dispersion dominates the signal by orders of magnitude, i.e. $\sigma_\gamma \ll \sigma_\epsilon$. For FRBs, however:
\begin{equation}
    \sigma_\mathcal{D}^2 =\int\frac{\ell\mathrm{d}\ell}{2\pi}C^{\mathcal{D}\mathcal{D}}(\ell)\;,
\end{equation}
is significant due to the strength of the signal. Additionally, the host galaxy contribution acts as a shot noise contribution with variance $\sigma^2_\mathrm{host}$ per event, 
thus adding a white noise contribution to the observed spectra:
\begin{equation}
\label{eq:dispersion_measure_angular_power_spectrum_observed}
    C^{\mathcal{D}\mathcal{D}}(\ell) \;\to\; C^{\mathcal{D}\mathcal{D}}(\ell)
    + \frac{\sigma_\mathcal{D}^2+ \sigma^2_\mathrm{host}}{\bar{n}}\;,
\end{equation}
\review{here $\sigma^2_\mathrm{host}$ should be seen as the effective host of the FRB sample
\begin{equation}
\label{eq:host_contribution}
    \sigma^2_\mathrm{host} = \int\mathrm{d}z\,n_\mathrm{FRB}(z) \frac{\sigma^2_\mathrm{host,0}}{(1+z)^2}\,.
\end{equation}}}

\subsection{Power Spectra}
The angular power spectra in \Cref{eq:lensing_nonlimber} and \Cref{eq:dispersion_measure_angular_power_spectrum} are integrals over the matter power spectrum and the electron power spectrum. Cross-correlating the probes also requires the matter-electron power spectrum. We use \texttt{pyhmcode} \citep{mead_accurate_2015,mead_hydrodynamical_2020,troster_joint_2022} which was fitted to the \texttt{BAHAMAS} \citep{mccarthy_bahamas_2017} and \texttt{COSMO-OWLS} \citep{le_brun_towards_2014} simulation suites to reproduce the power spectra of matter, gas, stars and pressure assuming a halo model. The baryonic feedback in \texttt{pyhmcode} is controlled by a single parameter, the strength of AGN encapsulated in $T_\mathrm{AGN}$. This AGN temperature should be seen simply as a nuisance parameter with no physical correspondence in the real Universe.
Higher AGN temperatures lead to stronger suppression at large $k$ (and a rise again at even smaller scales). For current cosmological surveys, this description is usually flexible enough, since more extended models are likely to be prior-dominated and therefore offer too much flexibility. For future surveys, however, more flexible emulators might be needed \citep{angulo_bacco_2021,arico_simultaneous_2021,arico_bacco_2021}, allowing for more freedom at non-linear scales due to baryonic feedback.

\subsection{Statistics}
Full sky surveys measure $2\ell+1$ independent modes per multipole $\ell$. For any projected fields $f_i(\hat{\boldsymbol{x}})$, one would derive modes $f_{i,\ell m}$ from a spherical harmonic decomposition. Assuming that the likelihood for the set of modes $\{f_{i,\ell m}\}$ is a multivariate Gaussian with zero mean and covariance (or 2-point correlation) $\boldsymbol{C}$ with components
\begin{equation}
   ( \boldsymbol{C}_\ell)_{ij} = \langle f_{i,\ell m}f^*_{j,\ell^\prime m^\prime}\rangle \equiv \delta^\mathrm{K}_{mm^\prime}\delta^\mathrm{K}_{\ell\ell^\prime}C^{f_if_j}_\ell\;,
\end{equation}
the Fisher matrix is given by \citep{tegmark_karhunen-loeve_1997}:
\begin{equation}
\label{eq:fisher}
    F_{\alpha\beta} = f_\mathrm{sky}\sum_\ell\frac{2\ell +1}{2}\mathrm{tr}\left(\boldsymbol{C}^{-1}_\ell\partial_\alpha \boldsymbol{C}_\ell\boldsymbol{C}^{-1}_\ell\partial_\beta \boldsymbol{C}_\ell\right)\;.
\end{equation}
Here, Greek indices label parameters, and the observed sky fraction $f_\mathrm{sky}$ accounts for incomplete sky coverage. The inverse of the Fisher matrix $\boldsymbol{F}^{-1}$ yields the covariance matrix of the parameters and serves as a lower limit for obtainable errors given the survey settings by means of the Cram\'{e}r-Rao bound.

Since the derivative in \cref{eq:fisher} is calculated numerically, small differences can propagate into the matrix inversion to obtain possible constraints. Therefore, we check the stability of the derivative in \cref{app:fisher_matrix}.  Lastly, we are also calculating the signal-to-noise ratio (SNR) of the measurement, i.e. we assume that one wants to measure the amplitude, $A$, of the cosmological term in \cref{eq:observed_lensing_spectra,eq:dispersion_measure_angular_power_spectrum_observed}. Using \cref{eq:fisher} for $\boldsymbol{C} = A\boldsymbol{S}+\boldsymbol{N}$, one finds
\begin{equation}
\label{eq:snr}
    \Sigma^2(\leq\ell) = f_\mathrm{sky}\sum_{\ell^\prime}^\ell \frac{2\ell^\prime +1}{2}\mathrm{tr}\left(\boldsymbol{C}^{-1}_{\ell^\prime} \boldsymbol{S}_{\ell^\prime}\boldsymbol{C}^{-1}_{\ell^\prime} \boldsymbol{S}_{\ell^\prime}\right)\equiv \sum_{\ell^\prime}^\ell \mathcal{S}^2(\ell^\prime)\;.
\end{equation}
In the case here, the covariance takes the form:
\begin{equation}
\label{eq:covariance}
    \boldsymbol{C}_\ell = 
    \begin{pmatrix}
       \boldsymbol{C}^{\mathcal{D}\mathcal{D}}_\ell & \boldsymbol{C}^{\mathcal{D}\gamma}_\ell\\
       \boldsymbol{C}^{\gamma\mathcal{D}}_\ell & \boldsymbol{C}^{\gamma\gamma}_\ell
    \end{pmatrix}\;.
\end{equation}
Note that noise between the DM and the shear is uncorrelated, hence $\boldsymbol{C}^{\gamma\mathcal{D}}_\ell$ does not have a flat noise contribution. Lastly, \Cref{eq:fisher}, assumes that all probes are measured over the same footprint $f_\mathrm{sky}$. In order to account for different footprints, we modify the Fisher matrix as follows:
\begin{equation}
    \boldsymbol{F}^{\mathrm{FRB} + \mathrm{lensing}} \to f_\mathrm{sky,\mathrm{min}}  \boldsymbol{F}^{\mathrm{FRB}+\mathrm{lensing}}  + (f_\mathrm{sky,\mathrm{max}} -f_\mathrm{sky,\mathrm{min}})\boldsymbol{F}^{\mathrm{FRB}/\mathrm{lensing}}_\mathrm{max}\;
\end{equation}
where $f_\mathrm{sky,min}$ and $f_\mathrm{sky,max}$ identify the small and larger footprint of the two surveys, respectively and $\boldsymbol{F}^{\mathrm{FRB}/\mathrm{lensing}}_\mathrm{max}$ is the corresponding Fisher matrix of the larger footprint.
This replacement assumes that the footprints from which the angular power spectra are estimated overlap completely, making it equivalent to studying the power spectra estimators with a purely Gaussian covariance.

\section{Results}
\label{sec:results}

\begin{figure}
    \centering
    \includegraphics[width = .32\textwidth]{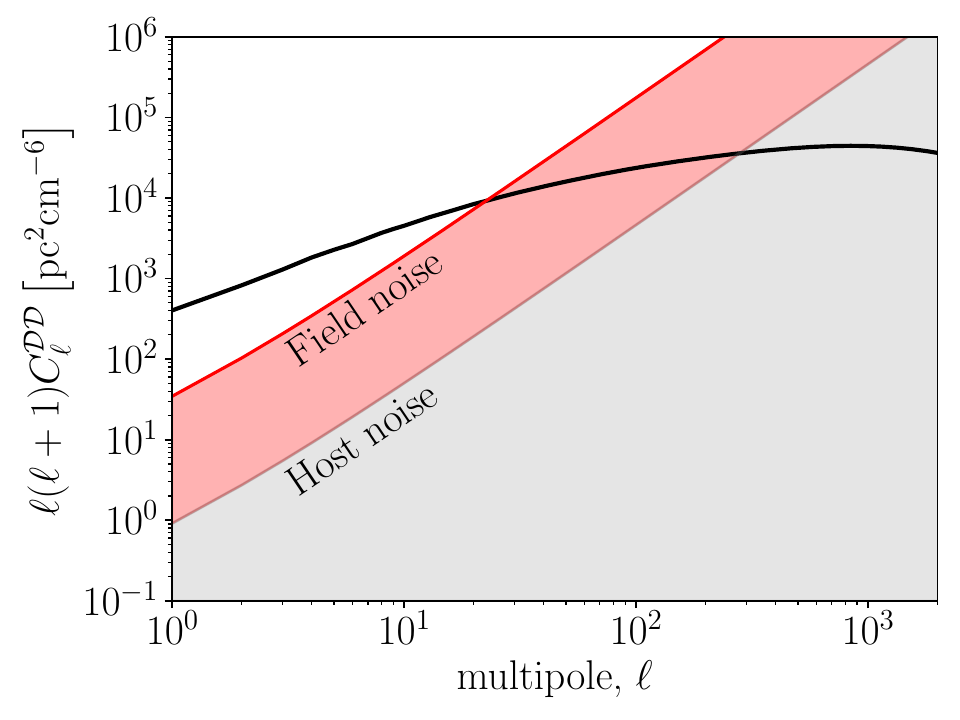}
    \includegraphics[width = .32\textwidth]{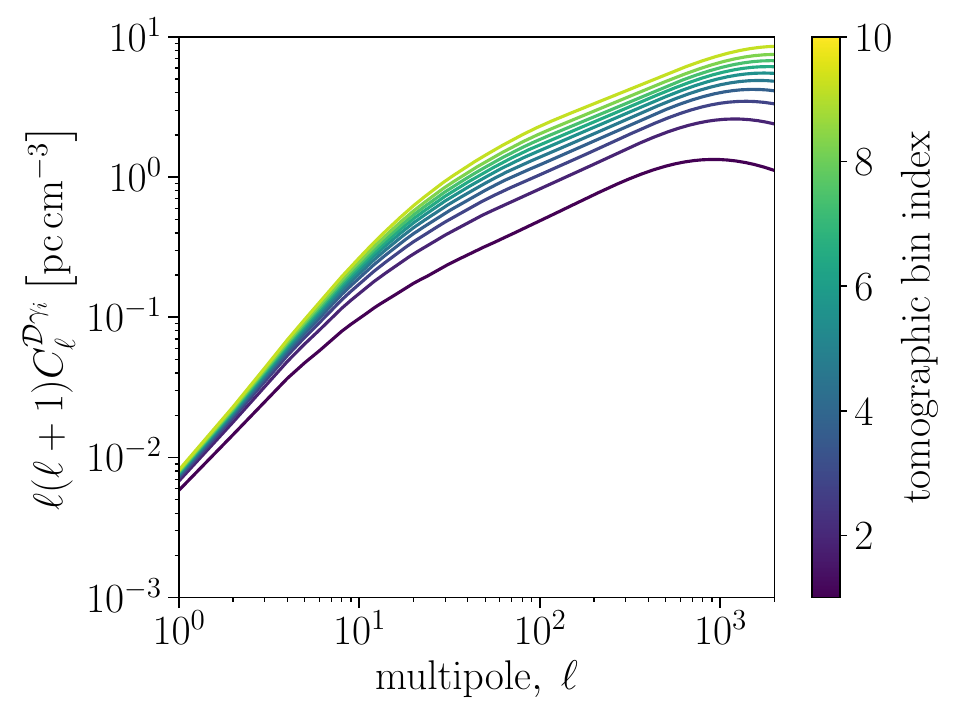}
    \includegraphics[width = .32\textwidth]{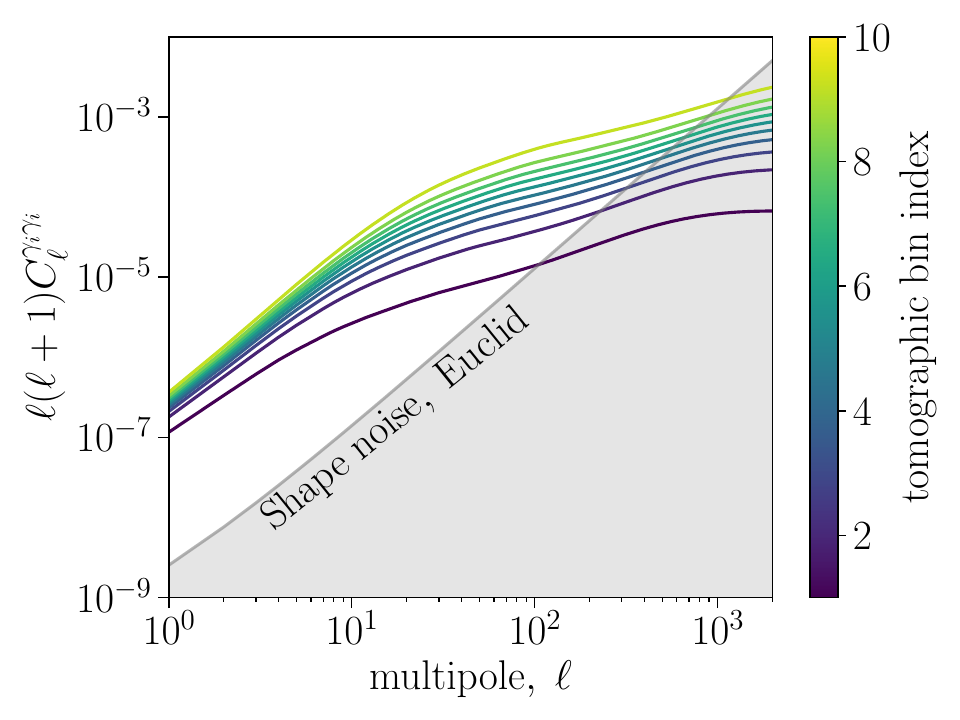}
    \caption{\reviewnew{Angular power spectra for the Euclid redshift distributions and an FRB sample with $\alpha = 3.5$ as defined in \Cref{eq:srd}. Note that all three plots have different units. The noise levels are, however, consistently scaled so that they represent the true noise level in each analysis. \textit{\textbf{Left}}: The solid black line shows the auto-correlation spectrum of the DM. The shaded grey and red areas indicate the DM noise arising from scatter in the host contribution and the intrinsic field noise for $N= 10^4$ FRBs, respectively. 
    \textit{\textbf{Centre}}: The dashed coloured lines show the cross-power spectra between the DM and the lensing convergence $\kappa_i$ in each of the ten tomographic Euclid redshift bins. The noise contribution for the cross-correlation vanishes. \textit{\textbf{Right}}: Auto-correlation, $i=j$, angular power spectra for the ten Euclid tomographic bins. The shaded region again quantifies the noise level for the Euclid cosmic shear. Note that we do not show the field noise here, as it is small compared to the shape noise contribution.}}
    \label{fig:cells}
\end{figure}

\subsection{Survey Specifications}
The survey specifications are given in \Cref{tab:survey_settings}, and the following redshift distribution for the FRBs is assumed
\begin{equation}
\label{eq:srd}
    n_\mathrm{FRB}(z) \propto z^2\mathrm{e}^{-z\alpha}\;,
\end{equation}
where $\alpha = 3.5$ controls the depth of the survey. For the weak lensing surveys, the redshift distributions are discussed in \citep{blanchard_euclid_2020} and \citep{2021A&A...647A.124H}. We show the weighting functions $W(\chi)$ for the individual surveys in \Cref{fig:weights}. The number of FRBs used to estimate the correlation is $5\times 10^4$, which is well within reach of the SKA \review{or DSA-2000 \citep{2019BAAS...51g.255H}, which is currently being built and aims for $> 10^4$ FRBs with localisation.}
The colour bar in \Cref{fig:weights} indicates the tomographic bin index for the weak lensing surveys. All weights are normalised to the maximum of the respective highest redshift bin. There is substantial overlap with Euclid (apart from the last tomographic bin) 

\review{\Cref{fig:cells} shows the three types of angular power spectra used in this analysis. On the left, the DM auto-correlation angular power spectrum is shown in black. \reviewnew{Furthermore, we indicate the intrinsic noise properties of this measurement by the red shaded area as calculated in \Cref{eq:dispersion_measure_angular_power_spectrum_observed}. From the figure, it is clear that the DM correlations dominate the intrinsic fluctuations due to the host contribution up to $\ell \sim 30$.
Above this multipole, the sum of the host noise and the field variance washes out the cosmologically interesting signal.} In the centre of \Cref{fig:cells}, we show the cross-power angular power spectrum between the DM and each of the ten tomographic cosmic shear power spectra as they will be measured by Euclid. Note that, since cosmic shear is dimensionless, the units of the spectra changed. In this case, there is no shaded area as the cross-spectra are noise-free. One can see that the curves are less flat at higher $\ell$ since the pure electron power spectrum is much more suppressed than the cross-power spectrum between matter and baryons at small scales. Lastly, we show the dimensionless cosmic shear angular power spectrum in the right panel. For more visibility, only the tomographic auto-correlation, $i=j$, is shown. Here, the steepness at larger $\ell$ is even more visible, again a result of fewer baryonic suppression of the total matter power spectrum compared to the electron power spectrum. The shaded area indicates the noise level of Euclid, which is the same in all tomographic bins as they are equi-populated and have the same intrinsic ellipticity dispersion, $\sigma_\epsilon$. \reviewnew{It should be noted that the field variance of the shear field is much smaller than the intrinsic ellipticity variance and is therefore not shown in the plot.}
}

\begin{figure}
    \centering
    \includegraphics[width = .49\textwidth]{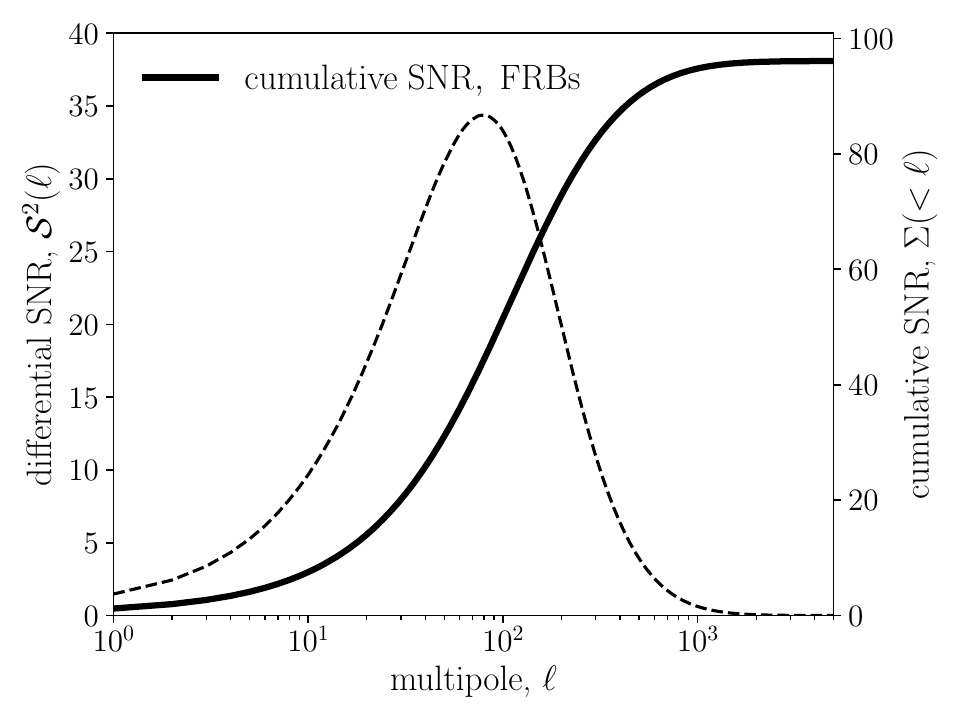}
    \includegraphics[width = .49\textwidth]{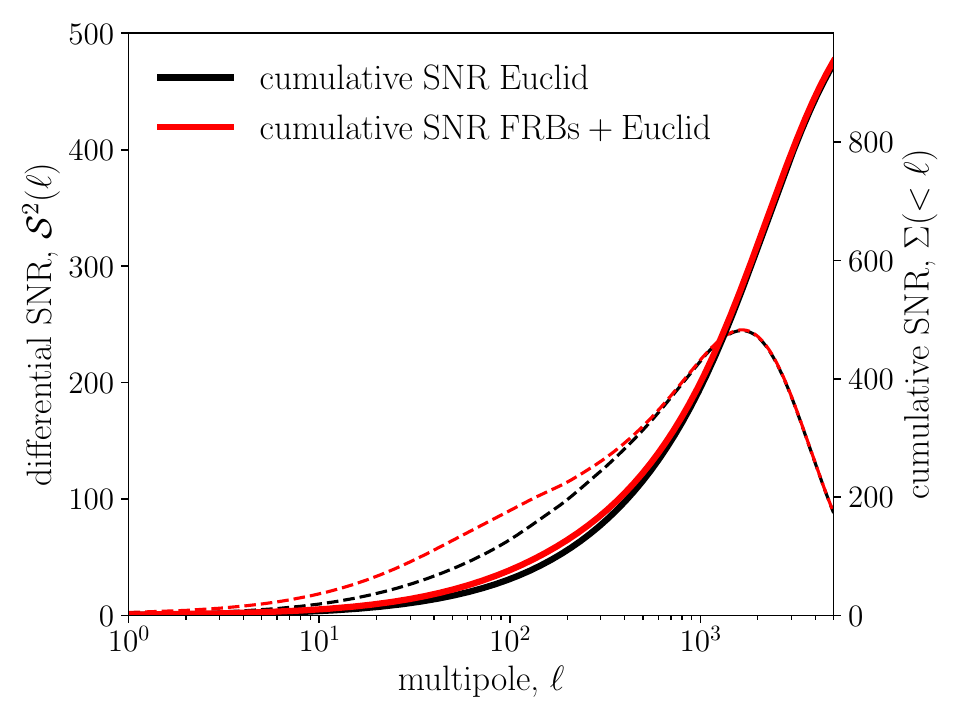}
    \caption{\reviewnew{Signal-to-Noise ratio (SNR) for a joint analysis of a cosmic shear survey with DM statistics. Dashed lines depict the squared SNR at each multipole, while solid lines show the cumulative SNR below the indicated multipole, $\ell$ (compare to Equation \ref{eq:snr}). {\textit{\textbf{Left}}}: for FRBs, {\textit{\textbf{Right}}}: for Euclid + FRBs. Note that the same FRB sample was considered in both cases. The number of FRBs in this plot is taken to be $10^5$.}}
    \label{fig:snr}
\end{figure}

\review{
Before discussing the potential SNR of the measurement, we would like to stress an important point. In \Cref{fig:cells}, the spectra have different dimensionality, since the corresponding fields have different dimensions themselves. Therefore, one should not compare the amplitude between the different $C_\mathrm{ell}$ directly. What is, however, important is the relative amplitude of each $C_\ell$ relative to the noise contribution, as they define the relative signal strength. As long as this is done consistently, it does not matter if one investigates relative fluctuations, as in the case of the cosmic shear, or dimensionful fluctuations, as for the DM. In this sense, the modes of the fields carry the correct statistical information, irrespective of the chosen unit system, provided their covariance is specified consistently. This is done by specifying the $C_{\ell}$ in \Cref{eq:covariance} at each multipole, $\ell$. This approach is equivalent to treating the estimated angular power spectrum as the model and concatenating it into a large data vector with a Gaussian covariance. Also, here, an entry in the data vector can have arbitrary and different units as long as this is reflected in the covariance as well. Since the pure noise contributions have the same dimensions as the $C_\ell$, this is taken into account properly in the present case. Therefore, we conclude that while comparing the power spectra directly to each other is not meaningful in any way, their SNR curves, as shown in \Cref{fig:snr} as a function of multipole $\ell$, are invariant under linear transformations (as unit transformations always are).} Dashed lines in \Cref{fig:snr} illustrate the squared differential SNR at each multipole, while solid lines show the integrated SNR. \reviewnew{The left plot shows the SNR for $5\times 10^4$ FRBs. We observe that the signal decays rapidly at higher multipoles due to field variance and host noise. An additional effect is the large $f_\mathrm{sky}$ in the case of the FRBs. The overall SNR is around 100. On the right, we show in black the Euclid-measured SNR for cosmic shear, reaching almost $10^3$, roughly ten times as high as that of the FRBs. The combined SNR is shown in red, where one can observe that the FRBs and their cross-correlations with cosmic shear add signal around $\ell =100$. It should be noted, though, that the red curve is not simply the sum of the two black curves from the left and the right plot due to the covariance between the different signals.}

\begin{figure}
    \centering
    \includegraphics[width = .7\textwidth]{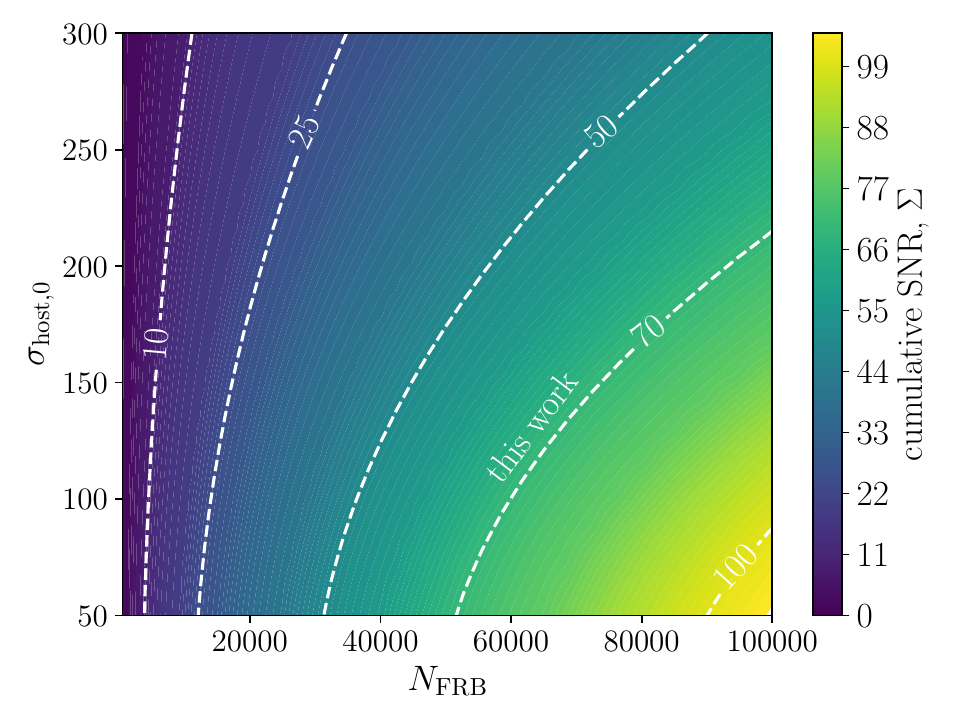}
    \caption{Cumulative SNR as a function of the Noise specification in \Cref{eq:dispersion_measure_angular_power_spectrum_observed}, where $\bar{n} = N_\mathrm{FRB}/(4\uppi)$. White lines indicate selected curves of constant SNR.}
    \label{fig:snr_function}
\end{figure}

\review{Reiterating on the previous discussion, we can employ a simple toy model to see whether the SNR ratios obtained are plausible by simply considering individual objects. The SNR for a single sheared lens in cosmic shear is roughly SNR$_\epsilon \approx 0.03$, assuming that the ellipticity dispersion is roughly 0.3 and the effect of cosmic shear on the ellipticity is roughly a per cent. For independent galaxy pairs, the total SNR of $N$ such galaxies is $\sqrt{N}$ SNR$_\epsilon$. Thus, for an SNR of $10^3$, one would require $\sim 10^9$ galaxies. We can play the same game with FRBs. Let us assume that the variance introduced by the host noise is of the same order as the one introduced by the LSS. This is a conservative estimate, especially at larger $z$, the LSS component will dominate. Therefore, the SNR of an individual FRB is unity, independent of whether we look at relative or absolute fluctuations. Hence, to achieve an SNR of 100 with $\sim 10^4$ FRBs. Therefore, the SNR we observe in \Cref{fig:snr} indeed make sense.  Since this is a very simple model and the DM scatter of the host galaxy, $\sigma_\mathrm{host}$, is quite uncertain, the total SNR is shown for different noise levels in }\Cref{fig:snr_function} where both $\sigma_\mathrm{host}$ and $N_\mathrm{FRB}$ are varied while the rest of the survey specifications is kept fixed (i.e. the distribution of FRBs along the line-of-sight). It should also be noted that the host contribution follows rather a log-normal than a Gaussian distribution \citep{zhang_intergalactic_2021,mo_dispersion_2022} so $\sigma_\mathrm{host}$ should be seen as the corresponding variance to that distribution, which is an accurate description as long as one is considering 2-point correlations only. \review{The fiducial value of $\sigma_\mathrm{DM} = 50$ is on the lower end of the plot, and it might be considerably larger. On the other hand, $N_\mathrm{FRB} = 10^4$ is a conservative estimate for the number of FRBs available at later data releases by Euclid or LSST-Rubin at the end of the decade. We would therefore argue that the SNR assumed here is in reasonable reach in that time span. Lastly, we would like to stress that $\sigma_\mathrm{host}$ is not the only noise component in the measurement. The total error is given by a sum of cosmic variance and the host noise contribution. This is exactly taken into account in \Cref{eq:fisher}, as $C_\ell$ includes both contributions.}

{
\begin{table}
\begin{center}
\tiny{
    \renewcommand{\arraystretch}{1.5}
    \begin{tabular}{cccccccccc}
        Probe & $h$  &$\Omega_\mathrm{cdm}$ &$\sigma_8$ &$\Omega_\mathrm{b}$  &$w_0$ &$w_1$ &\phantom{a}$\;
        \mathrm{log}_{10}T_\mathrm{AGN}$ &$\sum\! m_\nu [\mathrm{eV}]$ &$n_\mathrm{s}$\\ 
        \hline\hline
        &&&&\multicolumn{2}{c}{{\bf Fisher}}&&&&\\
        Euclid & 0.0168 & 0.0115 & 0.0167 & 0.0083 & 0.13 & 0.5705 & 0.0713 & 0.165 & 0.0388 \\
        FRB & 2.4556 & 2.5891 & 2.4621 & 0.2397 & 25.7362 & 183.219 & 1.8647 & 5.8782 & 3.5017  \\ 
        \hline
        Euclid+FRB & 0.0155 & 0.0102 & 0.0153 & 0.0014 & 0.118 & 0.4996 & 0.0139 & 0.1276 & 0.0272   \\
       & $( 8.0 \,\%)$ & $( 13.0 \,\%)$ & $( 9.0 \,\%)$ & $( 491.0 \,\%)$ & $( 10.0 \,\%)$ & $( 14 \,\%)$ & $( 411 \,\%)$ & $( 28 \,\%)$ & $( 43.0 \,\%)$   \\
        \hline\hline\\
        &&&&\multicolumn{2}{c}{{\bf MCMC}}&&&&\\
        Prior & $[0.64,0.82]$ & $[0.051,0.255]h^{-2}$ & $[0.7,1.2]$ & $[0.019,0.026]h^{-2}$ & $[-1,0]$ & $[-1 -w_0, -w_0]$ & none & $[0,\infty]$ & none \\
        \hline
        Euclid & 0.115 &0.025 &0.037 &0.021 &0.278 &1.166 &0.24 &0.357 &0.135   \\
Euclid + FRB & 0.071 &0.022 &0.033 &0.005 &0.252 &1.023 &0.048 &0.27 &0.074 \\
& $( 62.0 \;\%)$&$( 13.0 \;\%)$&$( 12.0 \;\%)$&$( 283.0 \;\%)$&$( 10.0 \;\%)$&$( 14.0 \;\%)$&$( 403.0 \;\%)$&$( 32.0 \;\%)$&$( 83.0 \;\%)$

\end{tabular}}  
\caption{One-dimensional marginal constraints for 68$\;\%$ confidence on the cosmological and feedback parameters using different survey settings. The first part of the table uses a Fisher analysis with no prior. In the lower part, we fit a noiseless data vector using MCMC and flat priors on some of the parameters.
For the Euclid+FRB case, we also quote the relative reduction in error as a percentage. \review{Note that the FRB case is only given for the full parameter set and uses $5 \times 10^4$ FRBs.}
}
    \label{tab:constraints}    
\end{center}
\end{table}
}

\begin{figure}
    \centering
    \includegraphics[width = .98\textwidth,trim={2.5cm 1cm 2.5cm 2cm},clip]{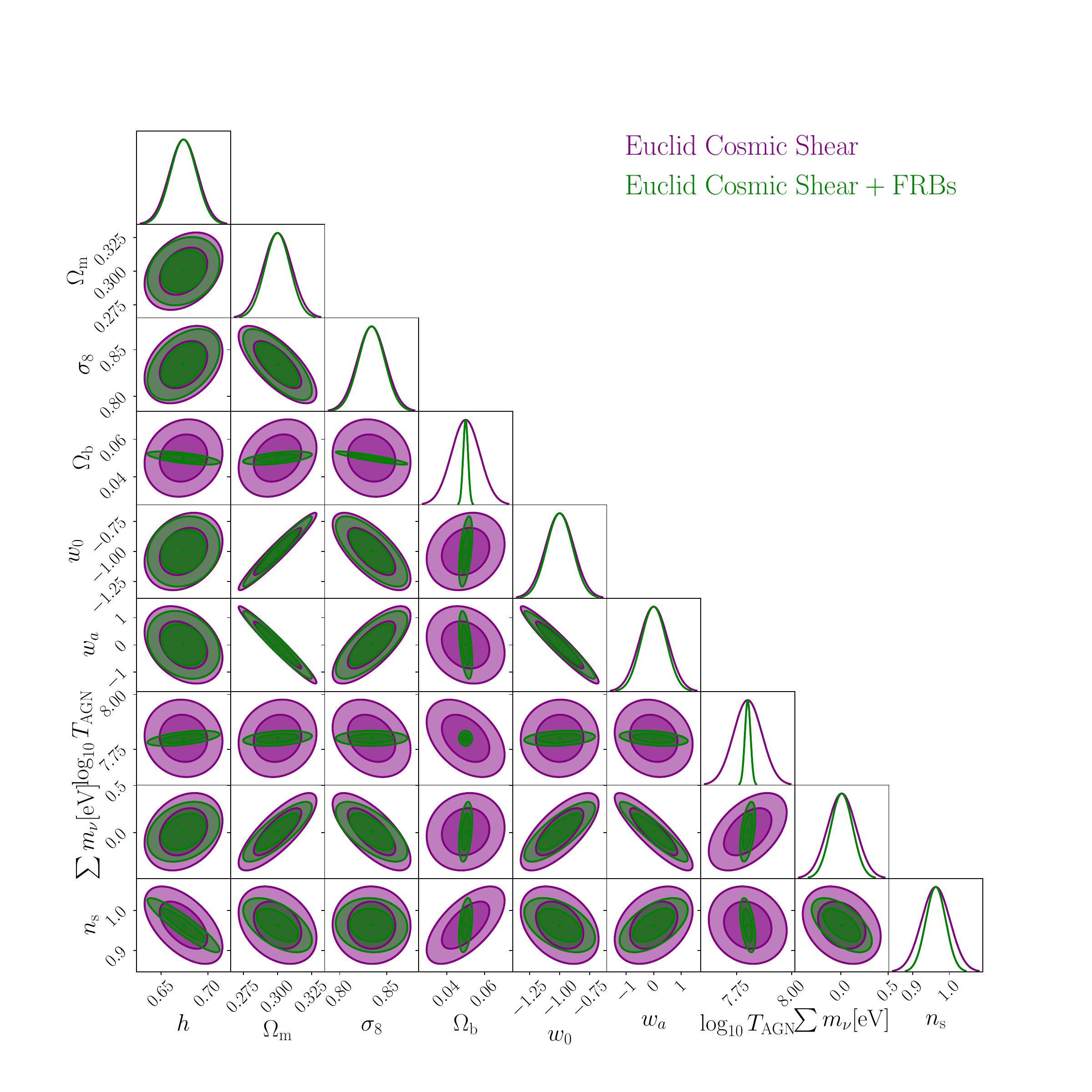}
    \caption{\reviewnew{Fisher matrix forecast for Euclid and an FRB survey with $N_\mathrm{FRB} =5\times 10^4$ and $\alpha = 3.5$. The contours indicate the 68 and 05 per cent confidence intervals. Blue contours show the constraints obtainable with, while green contours show the improvement when adding DM correlations. The footprints over which the observables are estimated are assumed to overlap completely. Contours from the FRBs are not shown in this figure, as they are outperformed for most parameters by cosmic shear in this case.}}
    \label{fig:fisher_euclid}
\end{figure}

\subsection{Cosmological Constraints}
We now analyse a set of 9 cosmological and feedback parameters for a stage-4, using the setup from the previous section and \Cref{tab:survey_settings}. The parameters and the corresponding one-dimensional marginal constraints are shown in \Cref{tab:constraints}. An analogous corner plot is shown in \Cref{fig:fisher_euclid} for the case of a stage-4 survey. In general, we distinguish two cases: $(i)$ FRB and cosmic shear auto-correlations separately and $(ii)$ FRB plus cosmic shear, also including the cross-correlation measurements. From \Cref{fig:snr} we already know that a stage-4 cosmic shear survey outperforms the considered FRB sample by almost a factor of 10 in SNR. It is thus expected that constraints on parameters that modify the amplitudes of both spectra simultaneously will be dominated by the cosmic shear measurement. This can be seen clearly in \Cref{fig:fisher_euclid}, where the amplitude of the power spectrum, $\sigma_8$, is very well constrained by Euclid. Similarly, one finds that the dark energy equation-of-state parameters $w_0$ and $w_1$ carry all the constraining power from the lensing signal as well. This is because the cosmic shear survey has a more extended redshift baseline and more tomographic information. The FRB sample, on the other hand, requires a lot of its signal to be put into constraining the total amplitude \cref{eq:DM_amplitude}, leaving almost no constraining power for the other parameters. \reviewnew{Note that this changes drastically when using FRBs with host identification \citep{walters_future_2018, hagstotz_new_2022} in a DM-$z$ analysis.} 

Typical parameters weakly constrained by lensing are the Hubble constant, $h$ and the baryon density $\Omega_\mathrm{b}$. Since both enter the DM amplitude, they can be well measured using FRBs. The following feature, however, is striking: the constraints on the feedback $\log_{10} T_\mathrm{AGN}$ are an order of magnitude better, including FRBs, compared to cosmic shear alone. The main driver of this effect is that the baryon distribution is much more affected by feedback than the total matter distribution, which consists mainly of dark matter, making it less responsive to feedback.
Similar performance can be expected for more complicated feedback models with more model parameters \citep{angulo_bacco_2021,arico_bacco_2021}. In fact, for more free parameters in the feedback, external constraints will become increasingly important. Due to the degeneracy between the sum of neutrino masses, $\sum m_\nu$, and the feedback parameter, $\log_{10}T_\mathrm{AGN}$, the former will benefit from the increased precision of the latter. Another interesting degeneracy is the primordial slope of the power spectrum, $n_\mathrm{s}$. Fixing this slope is essential to connect the clustering strength on small scales with the one measured on larger scales. Cosmic shear alone can achieve this only if feedback is constrained, showing that FRBs help break this degeneracy as well. Measuring $n_s$ precisely is very important for the measurement of scale-dependent growth on large scales, as is predicted in many theories of modified gravity.

\begin{figure}
    \centering
    \includegraphics[width = .98\textwidth]{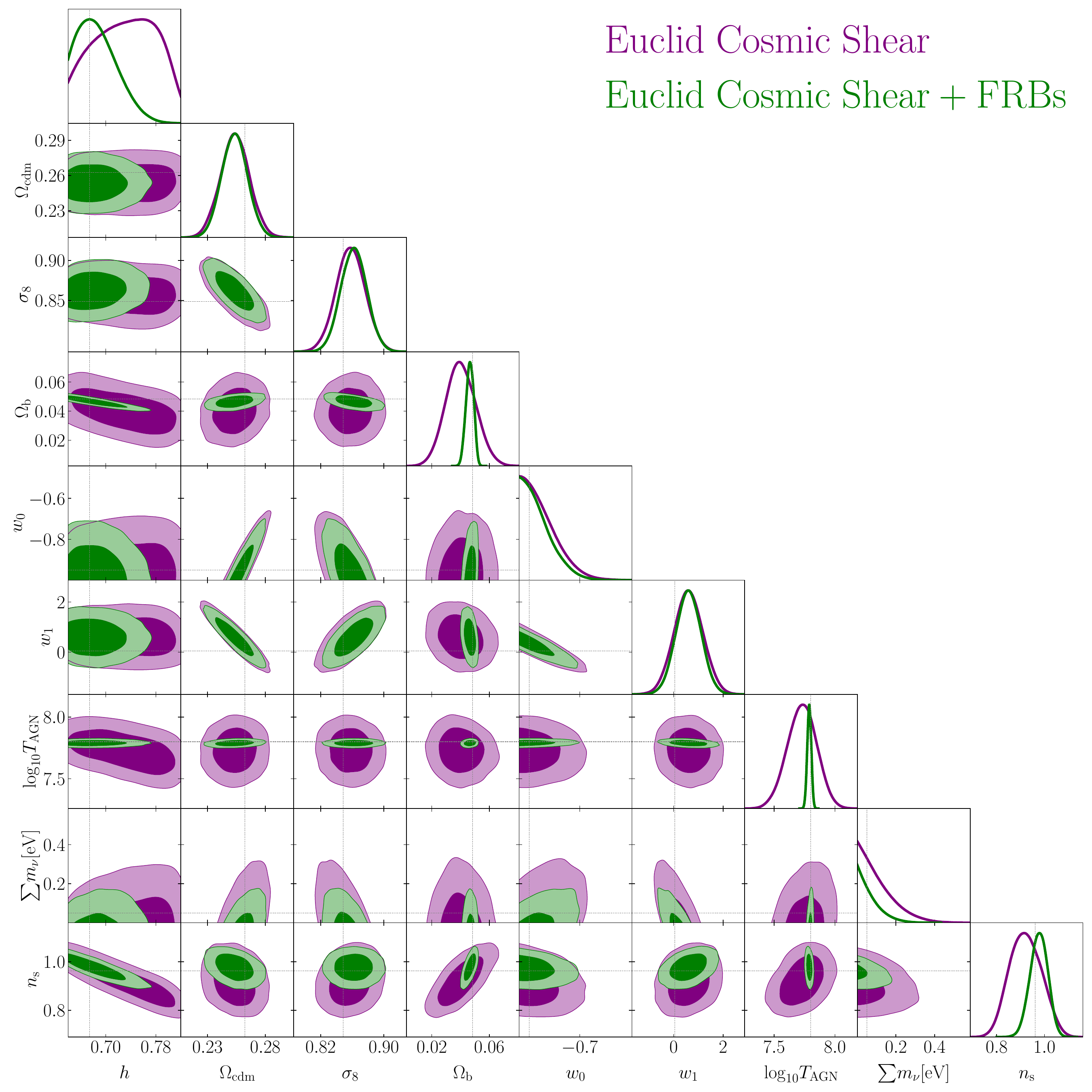}
    \caption{Forecast using a noiseless data vector for Euclid and an FRB survey with $N_\mathrm{FRB} = 5\times 10^4$ and $\alpha = 3.5$ (compare right panel of \Cref{fig:snr}). Line colours and styles are the same as in \Cref{fig:fisher_euclid} with the caveat that the priors from \Cref{tab:constraints} are used. The two contours correspond to the standard $68$ and $95\;\%$ confidence intervals.}
    \label{fig:contour_euclid}
\end{figure}

To mimic a more realistic situation, we create a noiseless data vector for the three spectra and estimate the posterior distribution using Markov Chain Monte Carlo (MCMC) sampled by \texttt{emcee} \citep{foreman-mackey_emcee_2013}. We impose some restrictions on the parameter space by imposing priors on the cosmological parameters. This particularly ensures that the nonlinear power spectrum and electron bias are not extrapolated beyond the range of simulations used to calibrate the model (see also table 1 in \citep{troster_joint_2022}). We also summarise the choices for the flat priors in the lower half of \Cref{tab:constraints} together with the marginal $68\%$ confidence intervals obtained by cosmic shear alone and cosmic shear combined with FRBs. The trends are very similar; however, we see less improvement overall. The reason is that the prior range provides additional information. It can be seen, for example, that the posteriors for $h$ or $\Omega_\mathrm{b}$ are prior-dominated in the cosmic shear case. Also, the restriction to non-phantom dark energies and no crossing of $w> 0$ reduces the parameter space volume. These effects reduce the gain from adding FRB measurements, since the prior already slightly breaks degeneracies. Nonetheless, we still find an order of magnitude improvement for the constraints on $\log_{10} T_\mathrm{AGN}$ and $50\%$  for the neutrino mass. \review{In \Cref{app:cl_fb} the effect of the cosmic shear angular power spectrum is shown, where it is clear that FRBs (green area) will substantially reduce the prior volume occupied by baryonic feedback models. To underscore the importance of this statement, it is worth noting that the exact functional form of the power spectrum suppression is unknown, and the case presented here is a minimal example. For more complex models with more parameters, the addition of additional probes becomes even more important.}

Finally, it is worth noting that the addition of FRBs also allows cosmic shear to serve as a more independent probe of the Cosmic Microwave Background (similar to using tSZ or kSZ), as external priors on the baryon density are not required, and the parameter can be fitted simultaneously.

\section{Conclusion}
\label{sec:conclusion}
\reviewnew{
The ability of future weak lensing surveys to address their key scientific questions, including the search for the mass scale of neutrinos, is largely limited by our understanding of the effect of baryonic physics on the clustering of matter on non-linear scales. While these effects can be measured from simulations, they are vastly different depending on the sub-grid model used. In this work, we investigated an alternative probe of the electron power spectrum, \review{which has been accessible only via the tSZ and kSZ effect so far}, the DM angular correlations measured from FRBs. DM correlations are an integrated quantity very similar to cosmic shear; they therefore provide access to scales comparable to those probed by cosmic shear and can efficiently break degeneracies between cosmological and feedback parameters. Compared to the tSZ and kSZ effects, DM correlations give different weights to different redshifts and scales. \review{While both these probes have, compared to DM correlations, already been detected, the addition of a new probe will achieve a couple of things. Of course, it will increase the available signal for LSS measurements. However, more importantly, it provides a completely new look at the baryon distribution in the Universe, which is independent of the CMB and has completely different systematics. Therefore, while still in its infancy, FRBs can offer an excellent alternative probe. It should also be noticed that FRBs, similar to the kSZ effect, test the electron density and are therefore potentially sensitive to similar physics. A cross-correlation between the three effects would therefore greatly check the different systematic aspects of the measurements.}
We used a simple feedback prescription with one parameter to model the matter, electron and matter-electron power spectra and investigated the improvements when cross-correlating a stage-4 cosmic shear survey with the DM from $5 \times 10^4$ FRBs, a number easily achievable within the next years. Our main findings can be summarised as follows:

\begin{enumerate}
\item[i)] The DM of FRB observations can improve constraints on baryonic feedback by an order of magnitude, depending on the SNR of the DM-correlation detection. Simultaneously, FRBs also help to constrain the Hubble constant and the baryon density. Combining these effects, the constraints on the neutrino mass can be improved by 25 to 50 per cent. The SNR used in this work is well within reach within the next decade, considering upcoming observations and dedicated searches for FRBs.

\item[ii)] DM correlations have a different radial weighting function than other baryon measurements, such as the SZ effect or X-ray measurements; they are therefore sensitive to different redshifts. Furthermore, the observations do not rely directly on the CMB and are hence an entirely different probe at low redshifts with different systematics. 
\end{enumerate}
\newreview{
There are a few limitations to our analysis. These include a very simple feedback model, Gaussian statistics and the absence of systematic effects. We also assumed that all FRBs have a host identification. While this is a target for surveys like DSA-2000 or CHORD, one can also consider a subset with known redshifts, which will help constrain the DM$-z$ relation and therefore allow the construction of an $n(z)$ for the whole sample.} 
\newreview{We show how this can be, in principle, done in \Cref{app:redshifts}, but save a detailed analysis of this for future work.} \review{The scope of this paper was to show that FRBs can be a powerful cosmological probe, in particular in conjunction with tracers of the full matter distribution, leveraging the different clustering properties of baryons and dark matter. In the future, we intend to address the limitations mentioned by adopting a more realistic inference setup. }
}

\acknowledgments
The authors would like to thank two anonymous referees for instructive comments.
RR is supported by the European Research Council (Grant No. 770935). SH was supported by the Excellence Cluster ORIGINS which is funded by the Deutsche Forschungsgemeinschaft (DFG, German Research Foundation) under Germany’s Excellence Strategy - EXC-2094 - 390783311.

\newpage
\bibliographystyle{JHEP}
\bibliography{MyLibrary,more_bib}

\newpage
\appendix

\begin{figure}
    \centering
    \includegraphics[width = .98\textwidth,trim={2.5cm 1cm 2.5cm 2cm},clip]{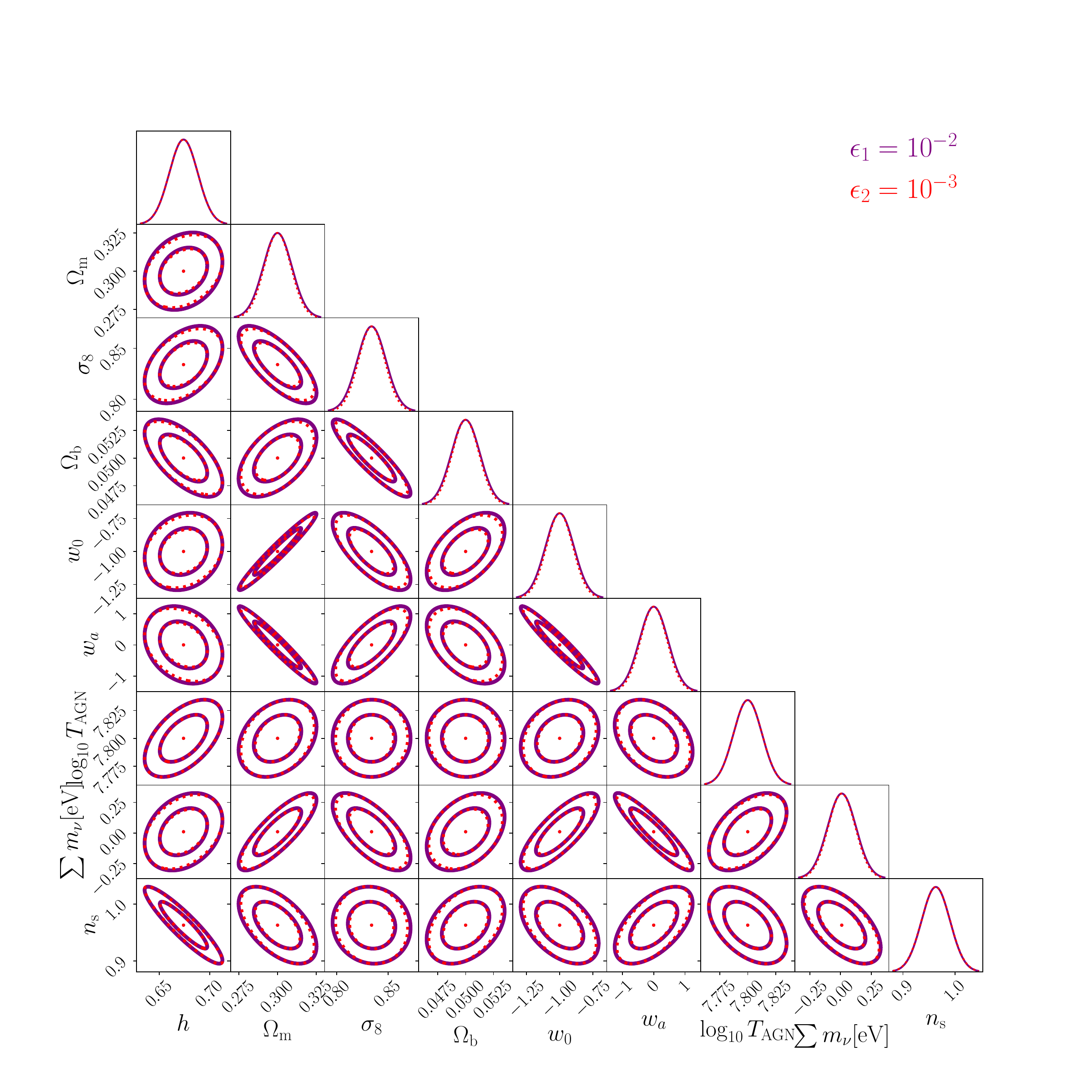}
    \caption{Same as \Cref{fig:fisher_euclid} but with varying step size for the numerical derivative. In particular, we choose $\epsilon_2 = 10^{-3}$ and $\epsilon_1= 10^{-2}$ for the green and dark violet lines, respectively. The fiducial case in \Cref{fig:fisher_euclid} assumes $\epsilon = 5\times 10^{-3}$.}
    \label{fig:fisher_euclid_eps}
\end{figure}

\section{Stability of the Fisher matrix}
\label{app:fisher_matrix}
To ensure the stability of the Fisher matrix, we vary the step size with which we evaluate the derivative. This exercise is shown in \Cref{fig:fisher_euclid_eps} for two different choices for $\epsilon$. In particular, we take the central form of the finite difference and vary the parameters by
\begin{equation}
    \delta\theta_{i\pm} = \left\{
    \begin{array}{ll}
      \theta_{i,0}(1\pm\epsilon)   &\;\;\text{if }\;\; \theta_{i,0}\neq 0 \\
               \pm\epsilon  & \;\;\text{else }  
    \end{array}\right.\;,
\end{equation}
where $\theta_{i,0}$ is the parameter's fiducial value.

\section{Effects of feedback on angular power spectra}
\label{app:cl_fb}
\review{
In \Cref{fig:feedback_cell} we show the effect of adding FRBs to a cosmic shear measurement. The coloured lines show different feedback models, quantified by the $\log_{10}T_\mathrm{AGN}$ parameter, and the expected Euclid statistical error bars for the power spectrum. It is clear that cosmic shear alone would need to put a significant fraction of its signal into constraining the baryonic feedback model (i.e. $\log_{10}T_\mathrm{AGN}$). However, adding FRBs narrows the range of possible feedback models to the green band, significantly reducing uncertainty and allowing cosmic shear to place more of the signal on other cosmological parameters. 

\begin{figure}
    \centering
    \includegraphics[width = .8\textwidth]{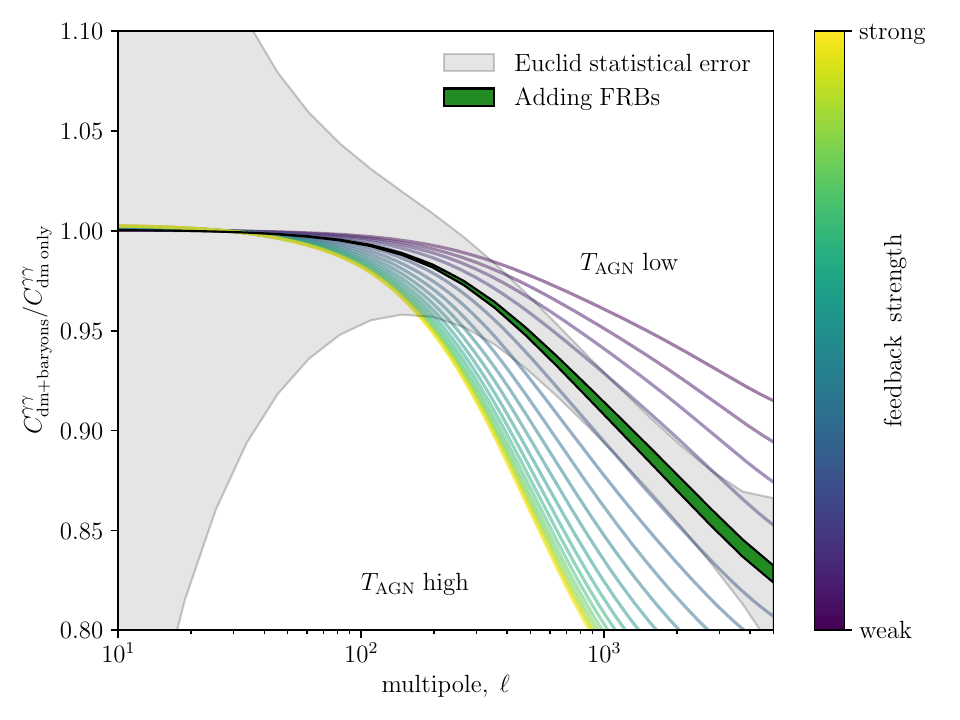}
    \caption{Effect of baryonic feedback on a typical angular power spectrum of cosmic shear. In particular, we show the ratio of the power spectrum with (dark matter plus baryons) and without (dark matter only) feedback. The colour bar shows the strength of feedback that scales with $T_\mathrm{AGN}$, giving an indication of the possible prior range of angular power spectra. 
    The grey band shows Euclid-like error bars around a reference feedback model with $\log_{10} T_\mathrm{AGN} = 7.6$. The green band shows by how much the feedback model will be constrained when FRBs are added to cosmic shear.}
    \label{fig:feedback_cell}
\end{figure}
}

\section{FRB redshifts}
\label{app:redshifts}
\begin{figure}
    \centering
    \includegraphics[width=0.8\linewidth]{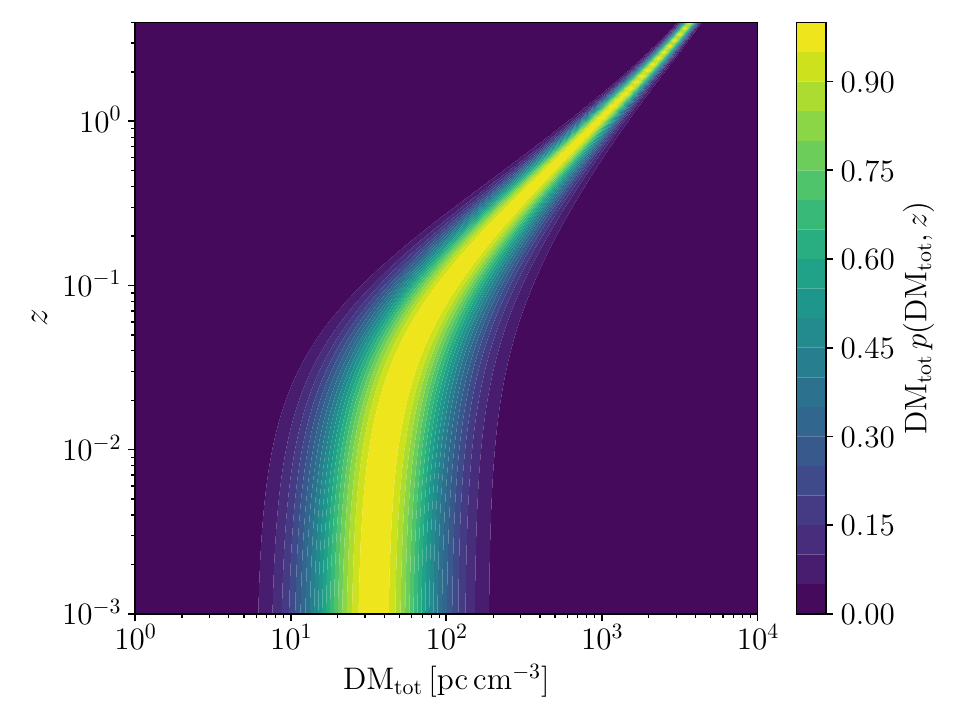}
    \caption{\reviewnew{Probability distribution function of the observed DM, $\mathrm{DM}_\mathrm{tot}$ as a function of $\mathrm{DM}_\mathrm{tot}$ and redshift. We normalise the probability distribution function such that it is unity at the peak for better visualisation.}}
    \label{fig:dmzpdf}
\end{figure}

In our analysis, we have assumed that all FRBs have an associated host galaxy and hence know redshifts. This allowed us to write down $n_\mathrm{FRB}(z)$ and obtain unbiased predictions for the statistical properties of the DM field. In reality, however, many FRBs may lack an associated host and therefore do not carry intrinsic redshift information. Suppose an FRB is observed at position $\hat{\boldsymbol{x}}$ without host identification with measured $\mathrm{DM}_\mathrm{tot}(\hat{\boldsymbol{x}})$, where we neglected the redshift dependence purposefully as it is not known \textit{a priori}. However, in principle, we can always invert the homogeneous DM-$z$ relation to obtain a noisy estimate of the redshift
\begin{equation}
\label{eq:redshift_estimate}
    \hat{z} \sim p_{\hat{z}}(z|\mathrm{DM}_\mathrm{tot}(\hat{\boldsymbol{x}}))\,,
\end{equation}
where we estimate $\hat{z}$ such that
\begin{equation}
\label{eq:est_redshift}
    \hat{z} =z\big(\langle\mathrm{DM}_\mathrm{LSS}\rangle\big)= z\big(\mathrm{DM}_\mathrm{tot}(\hat{\boldsymbol{x}}) - \langle\mathrm{DM}_\mathrm{host}\rangle - \mathrm{DM}_\mathrm{MW}(\hat{\boldsymbol{x}})\big)\,.
\end{equation}
As in \cite{reischke_probing_2021}, the redshift distribution, $n(z)$, is essentially given by marginalising over all possible observed DM:
\begin{equation}
    n(z) = \int \mathrm{d}\hat{z}\; n( \hat{z} ) p_{\hat{z}}(z|\mathrm{DM}_\mathrm{tot}) \,.
\end{equation}
The problem is now that for an individual FRB, the DM-$z$ relation is very noisy. This issue is depicted in \Cref{fig:dmzpdf} where we show the probability distribution function of the observed dispersion measure (compare Equation \ref{eq:dispersion_measure_contributions}). For this figure, we assume a lognormal distribution with mean $\langle\mathrm{DM}_\mathrm{tot}\rangle$ and variance $\sigma^2_{\mathrm{DM}_\mathrm{tot}}$:
\begin{equation}
\label{eq:sigma_unknown_redshift}
    \sigma^2_{\mathrm{DM}_\mathrm{tot}} = \sigma^2_{\mathrm{LSS}}(z) + \sigma^2_\mathrm{MW} + \sigma^2_\mathrm{host}(z)\;.
\end{equation}
Here $\sigma^2_{\mathrm{LSS}}(z)$ is calculated via the formalism discussed in \cite{reischke_covariance_2023}. We assume $\sigma^2_\mathrm{MW} = 30$.
Thus, given an observed DM, $\hat{z}$ has a large variance, or in other words, the DM field itself becomes much noisier. To illustrate this effect, we create Gaussian realisations of the DM field using an input $C^\mathcal{DD}_\ell$. We use \texttt{healpix} \citep{2005ApJ...622..759G,Zonca2019} with $N_\mathrm{side} = 1024$. The generated map is then evaluated at $5\times10^4$ randomly chosen pixels (FRBs) in the survey footprint (see \Cref{tab:survey_settings}). Each FRB is also associated with a true redshift from $n_\mathrm{FRB}(z)$. On top of the DM value from the LSS, we add the host noise and, in case noise from the unknown redshift according to the distribution shown in \Cref{fig:dmzpdf}. From the resolution realisations we measure the $C_\ell$ using \texttt{NaMaster} \citep{2019MNRAS.484.4127A,2025JCAP...01..028W}. The corresponding covariance is estimated directly from the realisations, with the Hartlap correction \citep{2007A&A...464..399H} applied.

\begin{figure}
    \centering
    \includegraphics[width = .95\textwidth]{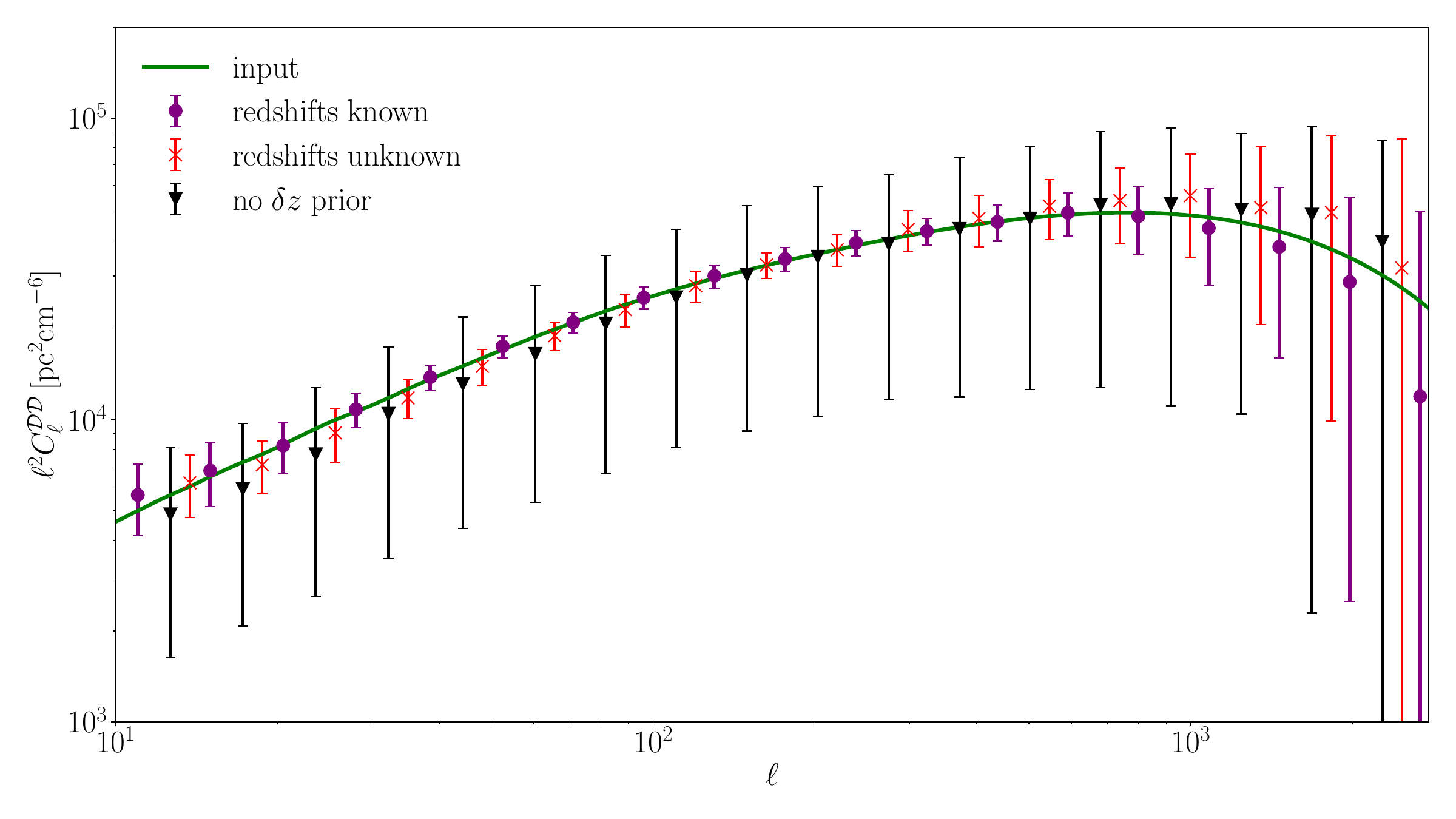}
    \caption{{Deconvolved angular power spectrum in 20 band powers estimated from $5\times 10^5$ FRBs. All bandpowers are estimated at the same multiples and are only shifted for better visualisation. The measurements show the mean of the realisations. Error bars indicate the uncertainty of the individual sample. The green line shows the input angular power spectrum used to generate the Gaussian realisations. Purple dots show the case where all redshifts are known, so that the catalogue only includes cosmic variance and host noise. This corresponds to the fiducial case studied in this work. Red crosses add additional noise to each FRB given by \Cref{eq:sigma_unknown_redshift}. Black triangles only use host noise but randomly shift the redshift distribution of FRBs according to \Cref{eq:redshiftshift}, introducing additional noise as well.}}
    \label{fig:autocalib_part1}
    \end{figure}

We run $100$ simulations and show the result in \Cref{fig:autocalib_part1}. Purple dots show the case that was discussed in the main part of the paper: redshifts are known, and only the host noise component is added (note that the field variance part in \Cref{eq:dispersion_measure_angular_power_spectrum_observed} is naturally taken care of in the realisations). The total SNR of the individual measurement from the 100 realisations is 68, as we would expect from \Cref{fig:snr_function}. As red crosses, we show the case where the redshifts are unknown. This substantially increases the uncertainties and reduces the overall SNR to 20. Note that the overall signal is always recovered because the estimator is insensitive to white noise. However, this adds to the uncertainties.

Lastly, we consider the following case: given an initial guess for the redshift distribution of the FRBs by means of \Cref{eq:est_redshift}, i.e. for each FRB there is a redshift estimate $\hat{z}$ and from this redshift estimate we can estimate the corresponding redshift distribution $n_{\hat{z}}(\hat{z})$. This redshift of each individual FRB will be extremely noisy, and the inferred distribution will also be subject to a lot of scatter. 
We model this residual noise by shifting the mean of the redshift distribution by some value $\delta z$, which is known to be sufficient for integrated effects such as cosmic shear or the DM \citep{2024MNRAS.530.4412R}:
\begin{equation}
\label{eq:redshiftshift}
    n_\mathrm{FRB}(z) \to  n_\mathrm{FRB}(z +\delta z)\,.
\end{equation}
For each realisation, we draw a random value of $\delta z\sim U(-1,1)$ and recalculate $C^\mathcal{DD}_\ell$. This introduces additional noise into the simulations, which is shown as the black triangles in \Cref{fig:autocalib_part1}. Again, the additional noise significantly increases the error bars. The overall SNR in this case is 2. To summarise, both unknown redshifts and $\delta z$ introduce significant noise in the catalogue, which masks the cosmological signal, whose amplitude we can only measure at $5\sigma$ or $2\sigma$ significance, respectively.

\begin{figure}
    \centering
    \includegraphics[width = .98\textwidth,trim={2cm 1cm 2.5cm 2cm},clip]{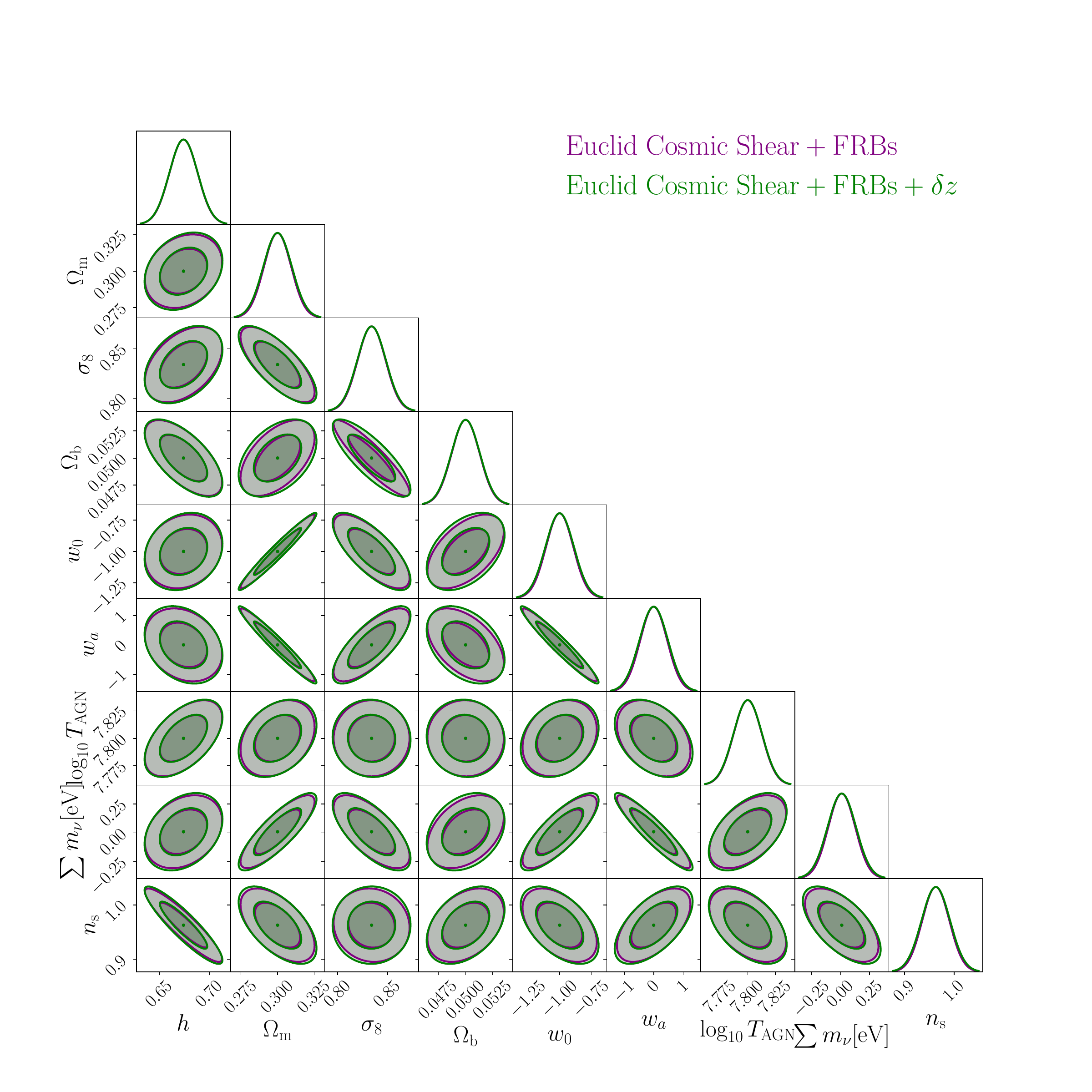}
    \caption{{Influence of marginalising over the redshift shift parameter $\delta z$ on the constraints on cosmological parameters. Green curves marginalise over $\delta z$ while the purple curves assume that it is known perfectly. Note that we only assume a shift parameter in the FRB redshift distribution. The source distribution for cosmic shear is perfectly known.}}
    \label{fig:redshiftshift}
    \end{figure}

Although the situation is not identical, this suggests that we can mimic the unknown redshifts of the FRBs by introducing the $\delta z$ parameter and marginalising over it in the analysis. The result is shown in \Cref{fig:redshiftshift} where the case with known redshifts is shown in black and with an unknown mean of the distribution in dashed (red). Clearly, the influence of the shift's marginalisation is minimal. The reason for this is that the redshift distribution can be self-calibrated very well because of the overlap with a calibrated redshift distribution of the cosmic shear sample. Although this analysis is highly simplified, it shows that, in principle, a substantial fraction of FRBs can be used to host identification for cross-correlation studies, as presented in this work. Another reason the influence of the unknown redshifts is small is that the effect investigated here has a complicated redshift and scale dependence, which allows us to distinguish between them.

\end{document}